\documentclass[twocolumn,trackchanges]{aastex631}

\newcommand{\cuo}[1]{\mbox{C$^{#1}$O}} 
\newcommand{\kms}{\hbox{km\,s$^{-1}$}}

\usepackage{enumitem}

\usepackage{amsmath}

\received{August 22, 2024}
\accepted{May 28, 2025}

\submitjournal{Astrophysical Journal}

\shorttitle{Infall and Accretion Shocks in Disks}
\shortauthors{Terebey et al.}

\begin{document}
\title{The Dynamics of Infall and Accretion Shocks in the Outer Disk}

\correspondingauthor{Susan Terebey}
\email{sterebe@calstatela.edu}

\author[0009-0002-9398-2500]{Susan Terebey}
\affiliation{Department of Physics \& Astronomy, California State University at Los Angeles, Los Angeles, CA 90032, USA}

\author[0000-0001-8568-8729]{Loraine Sandoval Ascencio}
\affiliation{Department of Physics \& Astronomy, University of California, Irvine, 4129 Reines Hall, Irvine, CA 92697, USA}

\author[0000-0001-8906-1528]{Lizxandra Flores-Rivera}
\affiliation{Centre for Star and Planet Formation, Globe Institute, University of  Copenhagen, \O ster Voldgade 5-7,   
1350 Copenhagen, Denmark}

\author[0000-0001-8292-1943]{Neal J.\ Turner}
\affiliation{Jet Propulsion Laboratory, California Institute of Technology, 4800 Oak Grove Drive, Pasadena, CA 91109, USA}

\author{Andrew Barajas}
\affiliation{Department of Physics \& Astronomy, California State University at Los Angeles, Los Angeles, CA 90032, USA}

\begin{abstract}
High-spatial-resolution observations of disks around young stars suggest planetary systems begin forming early, during the protostellar phase ($\le$ 1 Myr) when stars accrete most of their mass via infall from the surrounding cloud. During this era shocks are expected to be ubiquitous around the gaseous accretion disk due to supersonic infall that strikes the disk. We investigate the role of shocks using a theoretical and modeling framework we call the shock twist-angle Keplerian (STAK) disk, connecting the disk and infalling envelope gas via a shock using general physical principles. Briefly, at the shock, energy is dissipated while angular momentum is conserved, so that the infalling gas must change direction sharply, yielding a bend or twist in the streamlines. The model's pre-shock gas follows free-fall parabolic trajectories, while the post-shock gas is on lower-energy, elliptical orbits.  

We construct synthetic observations and find that the deviations from circular Keplerian orbits are detectable in Doppler-shifted molecular spectral lines using radio interferometers such as ALMA. Specifically, the STAK model leads to line emission
intensity and velocity-moment maps that are asymmetric and offset with respect to the disk structure traced by the dust continuum. We examine archival ALMA data for the class 0/I protostar L1527 and find the \cuo{18} velocity moment map has features resembling the disk plus envelope emission that naturally arise when the two are connected by a shock.
Thus, spectral line observations having sub-\kms\ spectral resolution and angular resolution sufficient to fully resolve the disk can reveal protostars' envelope-disk shocks.

\end{abstract}



\section{Introduction} \label{sec:intro}

The classic theory of low-mass star formation known as inside-out collapse, hereafter TSC, follows the gravitational collapse of a slowly rotating, isothermal, and axisymmetric cloud to form a protostar plus disk, \citep{ Terebey_1984, Shu_1987}. The TSC collapse model has been widely used with Monte Carlo radiative transfer modeling codes including \citet{Whitney_2013}, to model the structure of protostellar systems and derive physical parameters based on spectral energy distributions and imaging observations, such as the HOPS survey of the Orion star-forming region \citep{Furlan_2016}. During the collapse phase, shocks are expected to occur where the infalling material enters the disk, but their presence is not typically included.

However, some protostars show complex motion possibly indicative of shocks that does not neatly fall into a standard paradigm of gas free-falling onto a disk, where the disk is modeled as an accretion disk or a disk having circular Keplerian orbits. The motivation for our study begins with strong evidence for shocks and non-circular motion in the outer disk of the Class 0/I protostar L1527 IRS (IRAS~04368+2557), for which \citet{Sakai_2014} and \citet{Oya_2015} present a self-described toy model for rotating and infalling gas, as traced by $cyclic$-C$_{3}$H$_{2}$ and CS, to explain several observed features in the kinematics. Of note, the red-shifted gas is not restricted to only one side of the disk minor axis, as expected for circular disk orbits, but instead extends into three of the four image quadrants; the blue-shifted gas mirrors this behavior (see our \S \ref{sec:velocity_sig} for additional context). Moreover, \citet{Sakai_2014} find that the SO emission is restricted to a ring-like structure at 100~au that is proposed to be a shock tracing the outer disk edge. \citet{Aso_2017} suggest a disk edge nearer to 85~au. Similarly, for the iconic disk-with-rings and gaps protostar HL Tau \citep{Alma_2015}, there is evidence of accretion shocks with the detection of SO and SO$_{2}$ spiraling inwards towards the center of the disk in the same region where the infalling gas is probed by HCO$^{+}$ \citep{Garufi_2022}. For the Class I source TMC1A, \citet{Aso_2021} present evidence for spiral-like structure within 150~au of the protostar.

Kinematic methods that are relevant for detecting shocks and noncircular motion are detailed in recent work that examines protoplanetary disks later in the star formation process (i.e. Class II/T Tauri) as a tool to find protoplanets; see review by \citet{Pinte_2023}. The method relies on modeling gas motions in the disk and looking for velocity discontinuities as a way to identify the locations of potential protoplanets. Such kinematic deviations are more easily seen in disks with low turbulent motions (i.e. low turbulent viscosity), which occur at later times when infalling envelopes have largely dispersed. These studies have shown that a protoplanet forming in the protoplanetary disk can produce kinematic signatures in the gas that can be observable using long baseline configurations with the ALMA interferometer. One of the main kinematic signatures is caused by the Lindblad resonances in the disk from the embedded protoplanet resulting in a perturbation of the local Keplerian velocity pattern \citep{Ogilvie_2002, Teague_2018, Pinte_2018, Dong_2019}. The structure of the perturbed Keplerian velocity profile is analyzed by measuring the rotation curves of CO and its isotopologues emission relative to circular Keplerian rotation, namely, using moment-1 velocity maps. However, numerous other physical mechanisms can produce deviation from the local Keplerian velocity, such as streamers at near free-fall velocities \citep{Casassus_2015}, accretion shocks in the outer disk \citep{Sakai_2014}, and radial flows or warps \citep{Walsh_2017}. 

To better study young systems with mass accretion, where gas flowing onto and through the disk surface is important, we develop a new premise for a dynamic disk, whose surface is defined by a shock boundary, and that generalizes the motion and orbits present within the disk to be non-circular. We refer to this motion as the shock twist angle Keplerian (STAK) disk. In the simplest case that is described here, infalling gas flows through a shock, where it abruptly turns, and settles into an elliptical orbit. 

To fix ideas and to place the assumed system in context, the structure throughout much of the early protostellar phase consists of a collapsing cloud core (10,000 au; also known as the infalling envelope), a protostellar disk (tens of au), and a jet/bipolar outflow. The embedded and mostly obscured object at the center is a star-like object, the protostar (0.01 au). The mass of the protostar grows over time as infalling and rotating gas accretes onto the protostar, mostly inwards through the disk. To distinguish between different accretion regions, we follow a convention that denotes gas in the envelope (i.e. inner dense core) as infalling, and gas in the disk as accreting. During Class 0 the protostar gains roughly half its mass \citep{Andre_2000}; during Class I it reaches its final mass and the infall envelope clears, so that the pre-main sequence star and disk are revealed in Class II. The STAK dynamic disk premise applies to the gas rich disks typically found during Class 0 and Class I.

The model advanced by \citet{Sakai_2014} to explain non-circular Keplerian motion in the outer disk region of L1527, considers gas motion in the mid-plane, of a collapsing cloud core, consistent with a standard assumption that the gas conserves angular momentum and energy. In free-fall the gas would follow a parabolic (ballistic) trajectory. In the standard picture, gas flowing inward along the equatorial mid-plane encounters a shock at or near the centrifugal (disk) radius R$_\mathrm{CR}$, the radius appropriate for a circular orbit given the angular momentum of the gas just reaching the disk edge. Viscosity and other disk processes can transport angular momentum outward and modify the disk edge. By contrast, \citet{Sakai_2014} propose the novel assumption that the gas continues to flow inwards, rather than being arrested at/near R$_\mathrm{CR}$; thus the gas achieves super Keplerian velocities while moving inwards to the centrifugal barrier R$_\mathrm{CB}$, located at one-half the centrifugal radius.

\citet{Jones_2022} further explore the importance of the centrifugal barrier during disk formation using a hydrodynamic numerical simulation. They perform a time-dependent calculation that includes a viscous shear term, and analyze force balance in the disk midplane. Their results do not confirm gas inflow as far inwards as 0.5~R$_\mathrm{CR}$ as proposed by \citet{Sakai_2014}; however they do find super-Keplerian velocities at $\sim0.9~\mathrm{R_{CR}} $, closer to the forming disk edge. In their simulation the ballistic flow is modified by both viscosity and gas pressure. Moreover, \citet{Jones_2022} suggest that disk expansion is constrained by the infalling cloud gas.

However, \citet{Sakai_2014} and \citet{Jones_2022} do not consider the importance of infalling gas that enters the disk from above and below the disk midplane. These streamlines are essential because they carry much of the infalling mass, and moreover they naturally contain lower angular momentum material, so that the gas seeks to move inwards after initially impacting the disk. 

To better match observations that show evidence for infall plus rotation, i.e. dynamic motion in the outer disk, we focus on including off-disk streamlines and show that a physically plausible solution exists. The gas flow occurs on a dynamic timescale (i.e. orbital period), and goes through a shock that naturally leads to Keplerian but non-circular orbits. The STAK disk may be thought of as a limiting case, of rapid motion with time scale on the order of one orbital period. No viscosity is included, but roughly speaking the motion corresponds to $\alpha \sim 1$ in an $\alpha$ viscous disk model; this correspondence is only meant to indicate that motion in the disk occurs on the orbital timescale. Implicitly we assume that a disk has formed, and that significant accretion is occurring. This is in contrast to treatments using a slower viscous timescale where nearly circular orbits would be expected, e.g. \citet{Cassen_1981, Furuya_2017}. The approach is also in contrast to long timescales such as in \citet{Shariff_2022}, who use an inviscid approximation to investigate changes in the outer disk boundary that occur on a $10^5$ year evolutionary timescale. 

In addition to using a dynamic timescale this approach also benefits from the simplicity of ballistic flow in gravity, similar to the ballistic model of \citet{Sakai_2014}. Although the focus is on protostars, the physics of gravity is general, and the formalism developed is likely to apply to other accreting systems. By design the flow is largely insensitive to the detailed physics that may be present in hydrodynamic simulations, in order to simplify and highlight the underlying dynamics.  Investigations using HD or MHD simulations to study infall-supplied disks include:  \citet{Hueso_2005}, \citet{Visser_2010}, \citet{Kuffmeier_2017}, \citet{Kuznetsova_2022}.

The STAK dynamic disk notably includes a physics-based prescription for incorporating shocks and adding regions off the midplane, giving a model of gas flow surrounding protostars that may be readily adopted in other studies, as an aid for studying and analyzing dynamical motion in the outer regions of gas rich disks. In \S \ref{sec:dynamicdisk} we describe the underlying assumptions and resulting orbits, along with a 2D axisymmetric physical picture. In this work, geometrically thin disks are assumed for simplicity. However, thick disks can be accommodated
using a straightforward extension of the method presented. To illustrate the use of the STAK dynamic disk, we provide results based on a standard TSC inside-out collapse.  \S \ref{sec:modifications_radchemt} describes modifications to RadChemT, a package that includes dynamics, radiative transfer, and astrochemistry. The results of the simulations are presented in \S \ref{sec:results},  highlighting new features predicted for velocity moment maps arising from elliptical orbits in the disk versus standard circular orbits. 

\begin{figure*}[ht!]
\includegraphics[width=8.9cm]{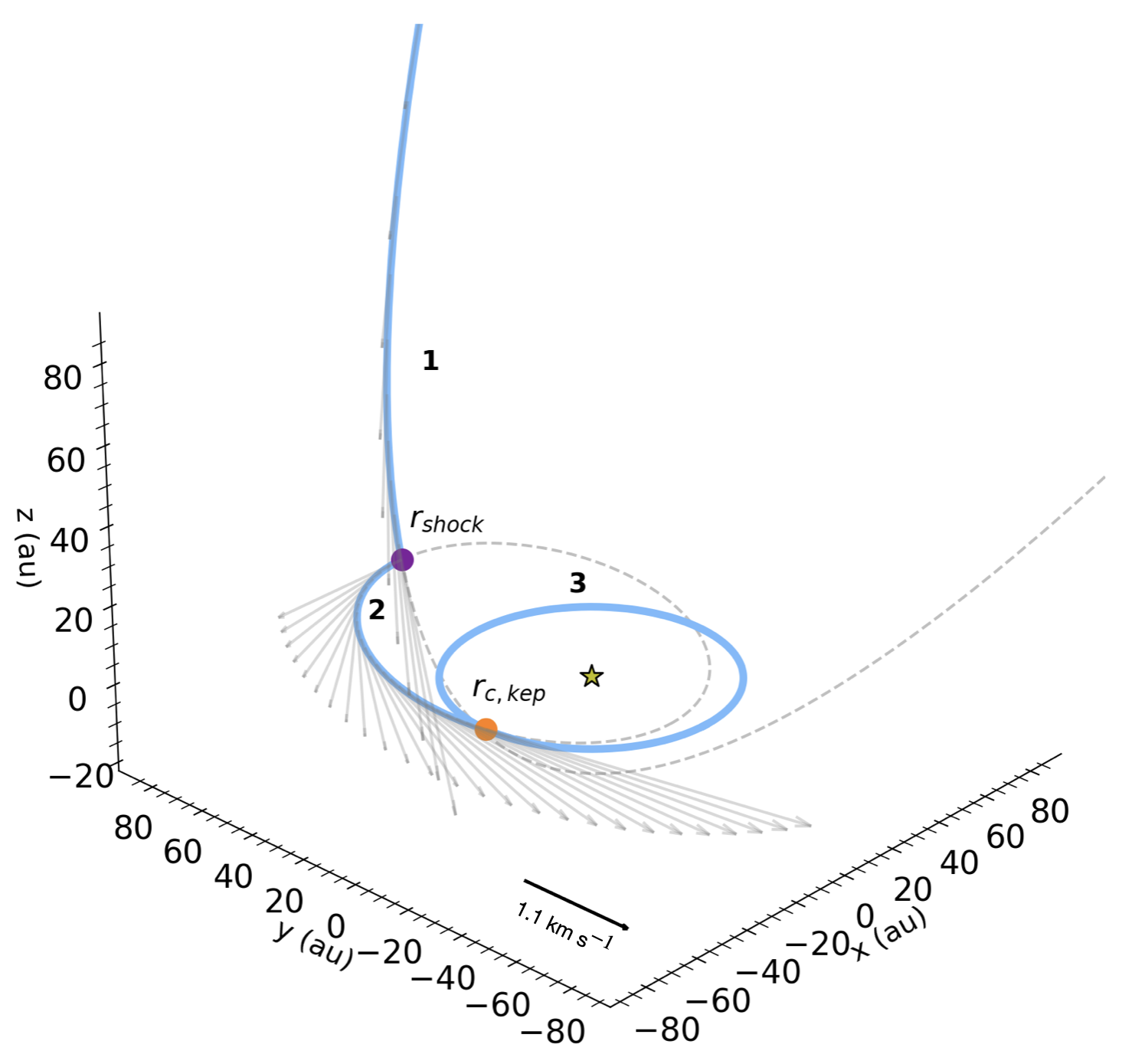}
\includegraphics[width=9.cm]{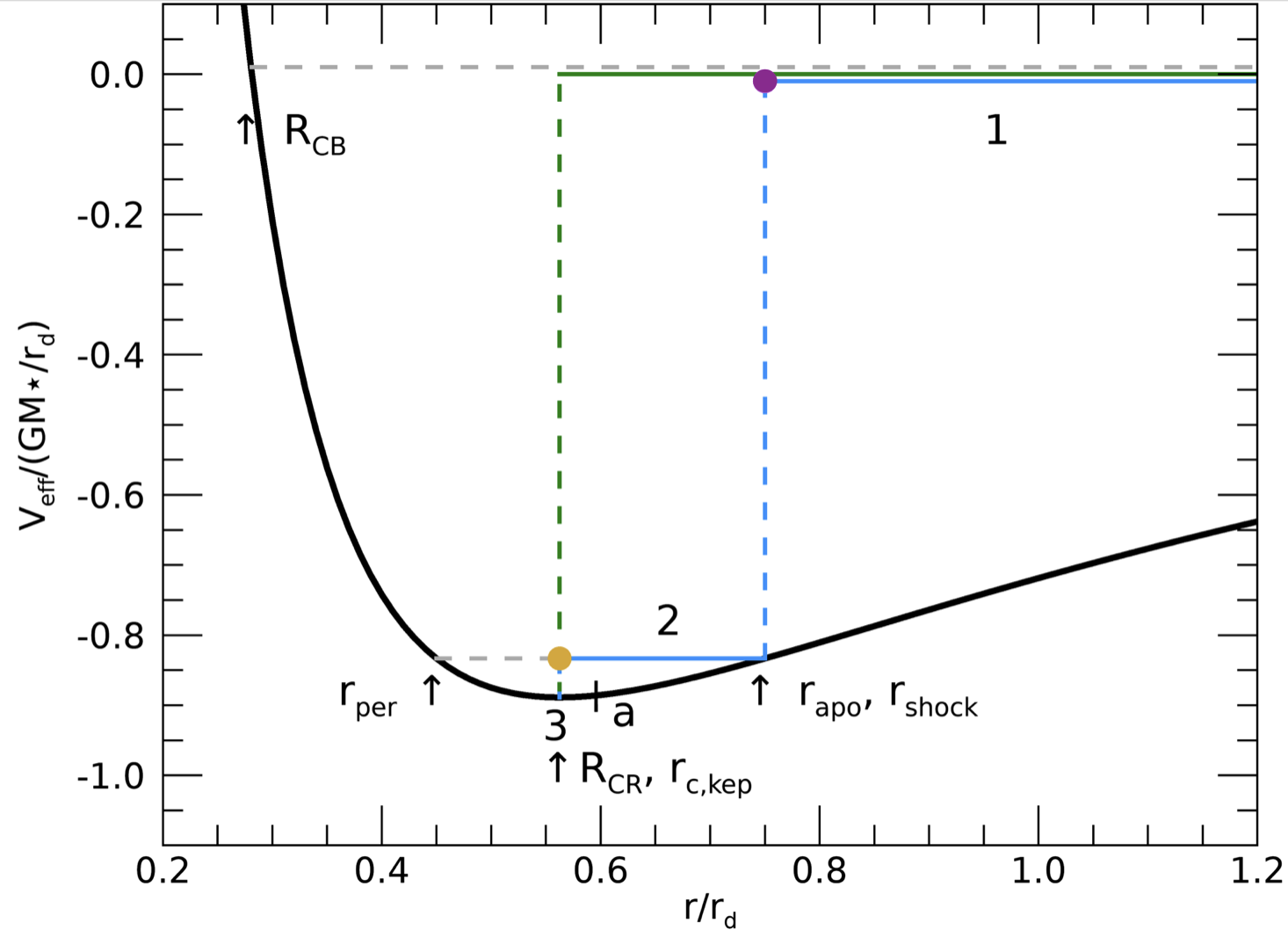} 

\caption{  \textit{Left} panel. Blue curve traces the path of a gas parcel, shown for the fiducial streamline (off-midplane, $\theta_{0}=60^\circ$) of the STAK dynamic disk. In region 1, envelope gas falls inward along a parabolic orbit. The gas encounters a shock (purple circle) when it enters the disk ($r_\mathrm{shock} < r_{disk}=r_{d}$), losing energy and transitioning to a lower energy elliptical orbit (region 2). The gas encounters a second shock (gold circle) when it hits the disk midplane, again losing energy and transitioning to a lower energy circular Keplerian orbit (region 3). Velocity vectors (in grey) are shown for the parabolic (region 1) and elliptical (region 2) but not circular (region 3) segments. \textit{Right} panel. Blue curve shows the corresponding effective potential for the fiducial (off-midplane at $\theta_{0}=60^\circ$) streamline of a STAK disk. Gas, conserving angular momentum, falls inwards (region 1) from infinity, in a zero energy parabolic orbit. If unimpeded, gas would reach the centrifugal barrier $R_{CB}$ (grey dashed curve). To attain a circular Keplerian orbit, at the appropriate centrifugal radius $R_{CR}=r_\mathrm{c,kep}$ (i.e. minimum effective potential, location 3) the gas must lose energy as illustrated by the green dashed curve. For the STAK disk, the gas parcel flows from the envelope (blue curve, region 1) across a strong shock at the disk surface (purple circle), losing energy but conserving angular momentum. The gas then follows the matching elliptical orbit inwards (region 2). At location 3, the gas crosses the disk midplane, where by necessity it undergoes a second energy dissipating shock (gold circle) and settles into a circular orbit having the minimum allowed energy.
\label{fig:effectivepotential}}
\end{figure*}

\section{DYNAMICAL GAS MOTION IN THE OUTER DISK} \label{sec:dynamicdisk}

This section presents the methodology of the STAK dynamic disk model, with the aim of changing the classical picture of collapse models to include shocks in the outer disk as a semi-analytical model. The focus of \S 2.1 is energy and angular momentum considerations, first describing the effective potential of a gas parcel traveling in parabolic motion from the envelope to the disk. Then, the fiducial case demonstrates how the gas parcel transitions from the parabolic envelope streamline to an elliptical streamline in the disk after experiencing the first shock at the disk surface.  In \S \ref{sec:parabolic_streamlines} and \S \ref{sec:elliptical_streamlines}, the focus is describing the velocity streamlines for the infall envelope, and the disk, respectively. Finally, \S \ref{sec:elliptical_streamlines}  together with \S \ref{sec:disk surface} shows how to determine the location of the disk-envelope shock boundary (i.e. disk surface) based on a ram pressure boundary condition; giving a new implementation of the physical modeling used in \citet{Flores-Rivera_2021}. 

\subsection{Potential energy prescription along streamlines}

Figure \ref{fig:effectivepotential} illustrates the basic premise of the STAK dynamic disk. The fiducial gas streamline traces a gas parcel falling in a gravitational potential, approaching from above the disk and having three distinct parts. Envelope gas starting far from the protostar falls inwards, tending towards a {\it parabolic} free-fall trajectory (region 1). At the disk boundary $r_\mathrm{shock}$ (purple dot), the gas parcel goes through an energy dissipating shock, transitioning from a parabolic to an {\it elliptical} orbit (region 2) within the disk. Notice the abrupt change in flow direction at the shock $r_\mathrm{shock}$ location. The gas parcel follows the elliptical orbit inward until it hits the disk midplane, where it goes through a second energy dissipating shock at $r_\mathrm{c,kep}$ (gold circle) and settles into a {\it circular} orbit (region 3). 

Physically, angular momentum is assumed to be conserved along the streamline. Energy is also conserved, except for dissipation at $r_\mathrm{shock}$ and $r_\mathrm{c,kep}$, the two shock locations. Under typical interstellar conditions, and in many collapse models, gas cools effectively. This means that pressure forces diminish in importance relative to gravity and the flow is supersonic approaching the disk (region 1), leading to free-fall parabolic streamlines. Encountering the disk as a ``brick wall'', the gas parcel goes through a strong oblique shock (i.e. flow velocity which is non-perpendicular to the shock front). The gas enters a vertically stratified disk, (i.e. horizontal density layers) so that the downward (vertical) velocity component will be quashed at the shock. Plausibly, and for simplicity, the inward (horizontal) velocity component is also quashed when meeting the stratified disk density layer. However, akin to a stone skipping on water, the azimuthal velocity $v_{\phi}$ can proceed unimpeded, thus $v_{\phi}$ is constant across the shock boundary. Therefore, post-shock, the gas parcel initially has no inward motion, but retains constant azimuthal $v_{\phi}$ motion, thus conserving its angular momentum at the shock location $(r_\mathrm{shock},\theta_{1},\phi_1)$ (see Table \ref{tab:parameters_description}). However, the force of gravity continues to act, and the gas parcel will therefore accelerate, while continuing to conserve angular momentum, thus transitioning to an elliptical orbit that is located within the disk. 

To aid in visualizing the gas streamline, recall in classical physics that an orbit is restricted to a two dimensional orbital plane. The angular momentum per unit mass $\Gamma_n$, normal to the orbital plane, is constant. The cylindrical component,

\begin{equation}\label{gamma_def}
\Gamma =\Gamma_n \sin(\theta_{0}), 
\end{equation}
aligned with the rotation axis of the cloud, is then also constant. For the fiducial streamline shown, the plane defining the parabolic section (region 1) is at polar angle $\theta_{0}=60^\circ$ with respect to the rotation axis. The plane defining the elliptical orbit (region 2) is at a different polar angle $\theta_{1}$, that is determined by the $r_\mathrm{shock}$ location given by a ram pressure boundary condition (see equation \ref{eq1} in \S \ref{sec:modifications_radchemt}). 

The polar angle of $\theta_{0}=60^{\circ}$ for the fiducial streamline is selected as characteristic in terms of mass infall from the envelope. A simple physical argument motivates this choice (TSC). Far from the disk at large $r$, the streamlines are approximately directed radially inwards. The mass infall rate is then roughly proportional to the relative solid angle $(\sim \Omega/4 \pi)$ coverage. The $\theta_{0}=60^{\circ}$ polar angle subtends half the total solid angle, so that approximately half of the mass entering the disk originates from streamlines at smaller polar angles, and half at larger polar angles. Table  \ref{tab:fiducial_parameters} provides additional parameters for the fiducial streamline.

To aid in understanding the energetics and angular momentum of the gas streamline, the effective potential (black solid line) in the \textit{right} panel of Figure \ref{fig:effectivepotential} shows possible orbits having different energy, but having the same angular momentum $\Gamma$. Recall that a constant energy orbit E appears as a horizontal line. The extrema, where the horizontal line touches the effective potential curve, define the minimum (periapsis) and maximum (apoapsis) radii of that orbit. For the fiducial streamline (blue line) gas falls inward along the parabolic streamline (region 1), which has by definition zero energy. At the disk boundary $r_\mathrm{shock}$ (purple dot) the gas loses energy across the shock as it enters the disk. The gas transitions to an elliptical orbit (region 2) at the new lower orbital energy, where $r_\mathrm{shock} = r_\mathrm{apo}$ becomes the apoapsis of the elliptical orbit (region 2). The shock location is identified as apoapsis because, as is well known for Keplerian orbits, the apoapsis turning point has nonzero azimuthal velocity but the radial velocity is momentarily zero; both conditions are consistent with our assumptions. Gas then moves inward along the elliptical orbit until it reaches the disk midplane at location 3 (gold dot) where it goes through a second energy dissipating shock, and settles into a circular orbit. The gas cannot cross the disk midplane, along the elliptical orbit, without encountering upwardly moving streamlines approaching the disk from below (dashed gray line from gold dot to $r_\mathrm{per}$ portion is therefore excluded). Thus at the disk midplane (location 3) occurs a second shock at $r_\mathrm{c,kep}$ and a region where material can accumulate (i.e. contact discontinuity). 

The effective potential curve in Figure \ref{fig:effectivepotential} is drawn for the $\theta_{0}=60^\circ$ fiducial streamline, and values have been normalized relative to $r_{d}$, the disk radius. Here we assume the disk radius is set by the streamline that carries the maximum angular momentum now entering the disk, which corresponds to the $\theta_{0}=90^\circ$ streamline. Other familiar points on the curve are the centrifugal barrier $R_{CB}$, namely the minimum allowed radius for a zero energy orbit, and the centrifugal radius $R_{CR}$, where a circular orbit occurs at the minimum allowed energy. For an axisymmetric system, rotation around the rotation axis generates a family of streamlines that all have the same angular momentum $\Gamma$ and same $r_\mathrm{shock}$ value, but differing $\phi_0$ azimuthal angles. Table \ref{tab:parameters_description} summarizes the coordinate system definitions.

As shown in \citet{Cassen_1981} for an infinitesimally thin disk, the parabolic streamline gas initially reaches the disk ($\theta=\pi/2$ in that work) moving at less than Keplerian velocity, namely $v_{\phi}/(GM/r)^{1/2} = \sin^2(\theta_{0}) \le 1,$ for $r \le r_d$. Therefore gas entering the disk at $r < r_d$ must lose energy and move inwards if it is to attain a circular Keplerian orbit.

It is convenient to characterize the orbit's angular momentum as \citet{Cassen_1981} do, in terms of where the parabolic orbit would cross the disk midplane if unimpeded by the disk, namely where $ \ell \equiv r $ at $\theta=\pi/2$ leading to,
\begin{equation}\label{ell_def}
\ell = \Gamma_{n}^2 / GM.
\end{equation}
Introducing a finite thickness disk, then a shock defines the envelope-disk boundary, and for that streamline one has $r_\mathrm{shock} \ge \ell$, since the shock is located upstream and above (or below) the midplane. 

The radius $r_\mathrm{c,kep}$ of the destination circular orbit can be calculated from the angular momentum of the parabolic streamline (region 1) as we show in Appendix \ref{app:ang_mom}, leading to
\begin{equation}\label{re_ckep}
r_\mathrm{c,kep}  = \ell  \sin^2(\theta_{0}),
\end{equation}
For the STAK dynamic disk, all three regions (1=parabolic, 2=elliptical, 3=circular) have the same cylindrical angular momentum $\Gamma =\Gamma_n \sin(\theta_{0})$. The effective potential connecting the three regions with constant $\Gamma$ is given by,
\begin{equation}
    V_{eff} = \frac{\Gamma^2}{2r^2\sin^2(\theta)} -\frac{GM}{r},
\end{equation}
and realizing that the expression for $V_{eff}$ simplifies when evaluated in the disk midplane at $\theta =\pi/2$, where $sin(\theta)=1$ and cylindrical $R=rsin(\theta)=r$. 
Moreover, considering the specific case that the precollapse cloud has constant angular rotation (solid body); this leads to,
\begin{eqnarray}
   \ell  = r_d \sin^2(\theta_{0}) \label{semilatus},\\
r_\mathrm{c,kep}  = r_d  \sin^4(\theta_{0}), 
\end{eqnarray}
See Appendices \ref{app:veff} and \ref{app:ang_mom} for the full derivation.

Notice that the STAK dynamic disk inherently does not assume vertical turbulent mixing at the location where gas enters the disk, and thus differs from standard alpha disk treatments. 

\begin{table*}[htp!]
\centering
\begin{tabular}{| c | c |}
\hline
Parameters & Description \\
\hline
$(r', \theta')$ & Plane polar coordinates of parabolic streamline\\
$(r', \alpha')$ & Plane polar coordinates of elliptical streamline  \\
$\theta_0$ & Orbital plane inclination of parabola with respect to z-axis \\
$\theta_1$ & Orbital plane inclination of ellipse with respect to z-axis \\
$r_{d}$ & Disk radius\\
$e$ & Eccentricity of elliptical streamline orbit  \\
$a$ & Semi-major axis of elliptical streamline orbit  \\
$r_{p}$ & Periapsis for the elliptical streamline orbit  \\
$r_\mathrm{shock}$ & Apoapsis for the elliptical streamline and shock location  \\
$r_\mathrm{c,kep}$ & Semi-latus rectum for the elliptical streamline and midplane point \\
$(r, \theta, \phi)$ & Spherical coordinates of the disk  \\
$\Phi$ & Line of nodes offset measured from the x-axis\\

\hline
\end{tabular}
\caption{Dynamical parameters utilized in the calculations of the parabolic and elliptical streamlines.}\label{tab:parameters_description}
\end{table*}%
 
\subsection{Parabolic gas streamlines within the cloud core}
\label{sec:parabolic_streamlines}

\begin{figure}[ht!]
\includegraphics[width=8.5cm]{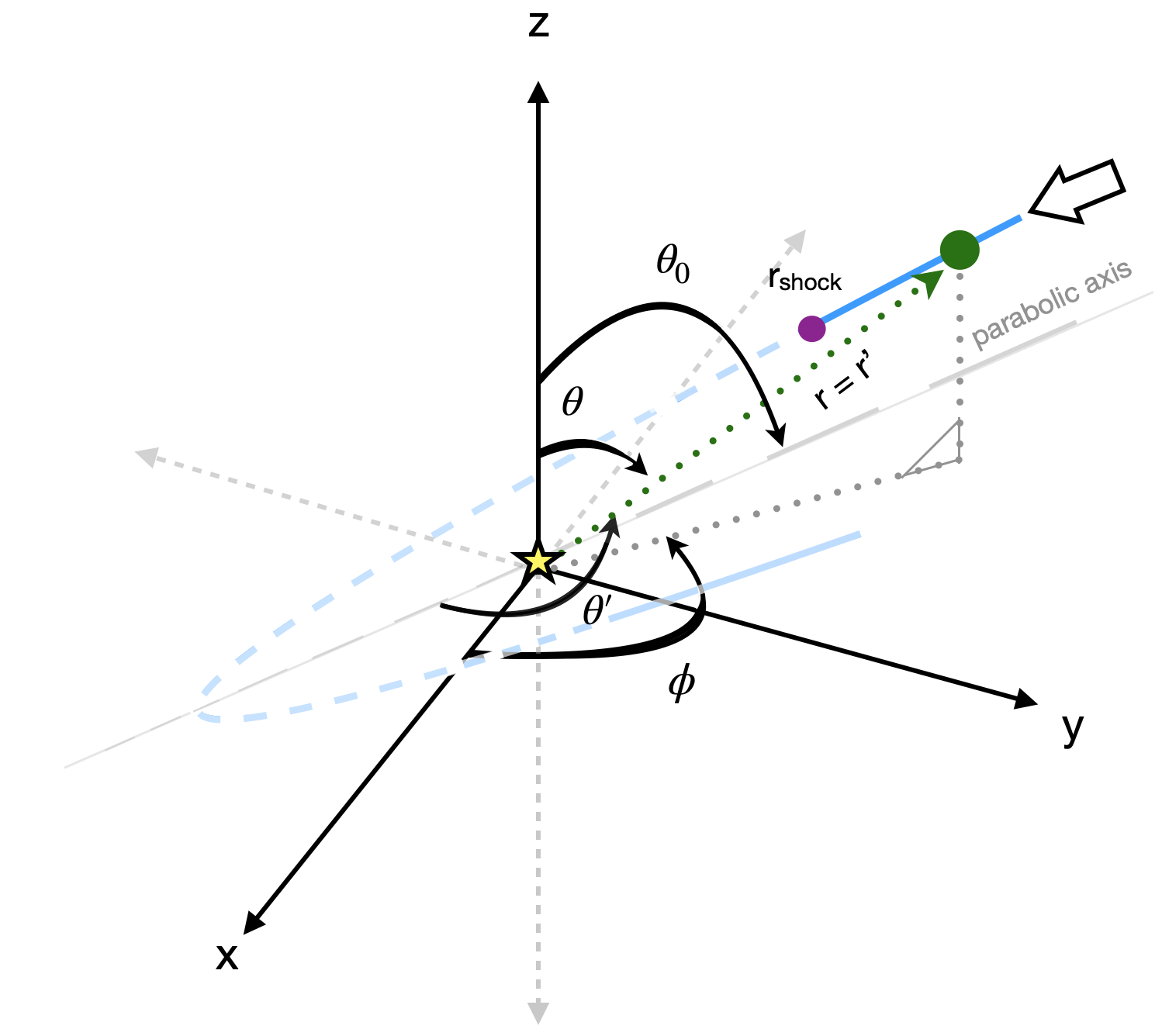}
\caption{Schematic of a parabolic streamline orbit of the gas parcel flowing from the envelope to the disk surface. The diagram is described in spherical coordinates (r, $\theta$, $\phi$). The gas parcel reaching at different infalling angles at the disk surface is described as plane polar coordinates (r$', \theta'$). The definition of the angles are specified in Table \ref{tab:parameters_description}.}
\label{fig:parabolicstreamline}
\end{figure}

We initially set up the problem as shown in Figure \ref{fig:parabolicstreamline}. We consider the protostar residing at the center of the coordinate system and the gas is rotating counterclockwise about the polar axis, defined by $\theta=0$ polar angle. 
For the infalling envelope, parabolic streamlines, representing the gas parcel reaching the disk surface (purple dot), are symmetric about $\theta=\pi/2$ and are also azimuthally symmetric. Figure \ref{fig:parabolicstreamline} describes a partial parabolic streamline such that gas parcels reach a shock once they reach the disk surface. 

This parabolic streamline prescription is established and discussed in \citet{Cassen_1981}. The angular momentum remains constant along a streamline and the assumption is that the disk mass is much smaller than the mass of the central protostar, leading to a conic orbit. The equation of the parabolic streamline orbit in plane polar coordinates ($r'$, $\theta'$) is described by
\begin{equation}\label{eq2}
r'= \frac{\ell}{1+\cos(\theta'),}
\end{equation}
where $\ell$ is the semi-latus rectum of the orbit, also referred to as the line of nodes. The radial value $r'=\ell$ occurs at $\theta'=\pi/2$; this location is where the parabola intersects the disk midplane ($z=0,\theta=\pi/2$). 
By letting $r'=r$, the angle definitions provide the transformation between the plane polar coordinates of the orbit and the spherical coordinates of the system,

\begin{equation}\label{eq3}
\cos(\theta')=-\sin(\phi)\frac{\sin(\theta)}{\sin(\theta_0)}, \qquad \sin(\phi)=\frac{\tan(\theta_0)}{\tan(\theta)}.
\end{equation}
This equation from \citet{Cassen_1981} defines an orientation for the parabolic streamline, so that both the line bisecting the parabola and an infalling gas parcel are directed inwards along the cartesian y-axis, and reach the disk midplane ($z=0$) with the line of nodes occurring along the cartesian x-axis. We adopt this convention for the reference parabolic orbit. If a different choice is desired then the orientation can be rotated by generalizing the argument $(\phi)$ to $(\phi-\phi_0)$ in equation \ref{eq3} for an arbitrary $\phi_0$ offset.

Combining equations (\ref{gamma_def}), (\ref{ell_def}), and (\ref{eq3}) and defining cylindrical angular momentum as $\Gamma = \Gamma_{\infty} f(\theta_{0})$, then equation \ref{eq2} can be re-written in spherical coordinates as 

\begin{equation}\label{eq4}
r=\frac{\Gamma_{\infty}^{2}}{GM} \frac{f^{2}(\theta_{0})}{\sin^{2}(\theta_{0})} \bigg(1-\frac{\cos(\theta)}{\cos(\theta_{0})} \bigg)^{-1}
\end{equation}
The equation relates the spherical coordinates ($r,\theta)$ of the parabola, given the polar angle $\theta_{0}$ that specifies the orbital plane of the streamline with respect to the z-axis. Parameter $G$ is the gravitational constant and $M$ is the combined mass of the protostar and the disk where $M$ (slowly) increases with time. For solid body rotation of the initial cloud, as we assume, the function $f(\theta_{0}) = \sin^{2}(\theta_{0})$. See Appendix \ref{app:ang_mom} for further discussion of how $\Gamma_{\infty}$ and the function $f (\theta_{0})$ relate to the angular momentum distribution of the precollapse cloud and to the collapse solution.

The functional form $\Gamma = \Gamma_{\infty}f(\theta_{0})$ describes the angular momentum distribution of the initial rotating cloud before collapse begins at $t=0$. The location where gas first hits the protostellar disk is dependent on this angular momentum distribution. To define the disk radius, we assume it equals the centrifugal radius corresponding to the maximum angular momentum now entering the disk, namely for the $\theta_{0}=90^\circ$ streamline where $f(\theta_{0})=1$, and therefore,

\begin{equation}\label{eq5}
r_d=\frac{\Gamma_{\infty}^{2}}{GM}.
\end{equation}

Equation \ref{eq4} can then be re-written in terms of the disk radius rather than the angular momentum as

\begin{equation}\label{eq6}
r=  r_{d} \sin^{2}(\theta_{0}) \bigg(1-\frac{\cos(\theta)}{\cos(\theta_{0})} \bigg)^{-1}.
\end{equation}

According to the collapse model (TSC or UCM), the velocity components of supersonic gas in the envelope are described by:
\begin{eqnarray}\label{eq7}
v_{r} &=& -\bigg(\frac{GM}{r} \bigg)^{1/2} \bigg(1+ \frac{\cos(\theta)}{\cos(\theta_{0})} \bigg)^{1/2} , \nonumber \\
v_{\theta} &=& \bigg(\frac{GM}{r} \bigg)^{1/2} \bigg(\frac{\cos(\theta_{0})-\cos(\theta)}{\sin(\theta)} \bigg) \bigg(1+ \frac{\cos(\theta)}{\cos(\theta_{0})} \bigg)^{1/2} \nonumber , \\
v_{\phi} &=& \bigg(\frac{GM}{r} \bigg)^{1/2} \bigg(\frac{\sin(\theta_{0})}{\sin(\theta)} \bigg) \bigg(1- \frac{\cos(\theta)}{\cos(\theta_{0})} \bigg)^{1/2}.
\end{eqnarray}
The magnitude of Equation \ref{eq7} is given by the free-fall velocity $v_{ff}=(\frac{2GM}{r})^{1/2}$.  Derivations for these velocity expressions are found in TSC, along with a description of how the UCM \citep{Ulrich_1976, Cassen_1981} and TSC solutions match at small $r$ in the infalling envelope. As noted in Appendix A of \citet{Shariff_2022}, some literature sources contain misprints for the velocity expressions.

\begin{figure}[ht!]
\includegraphics[width=8.5cm]{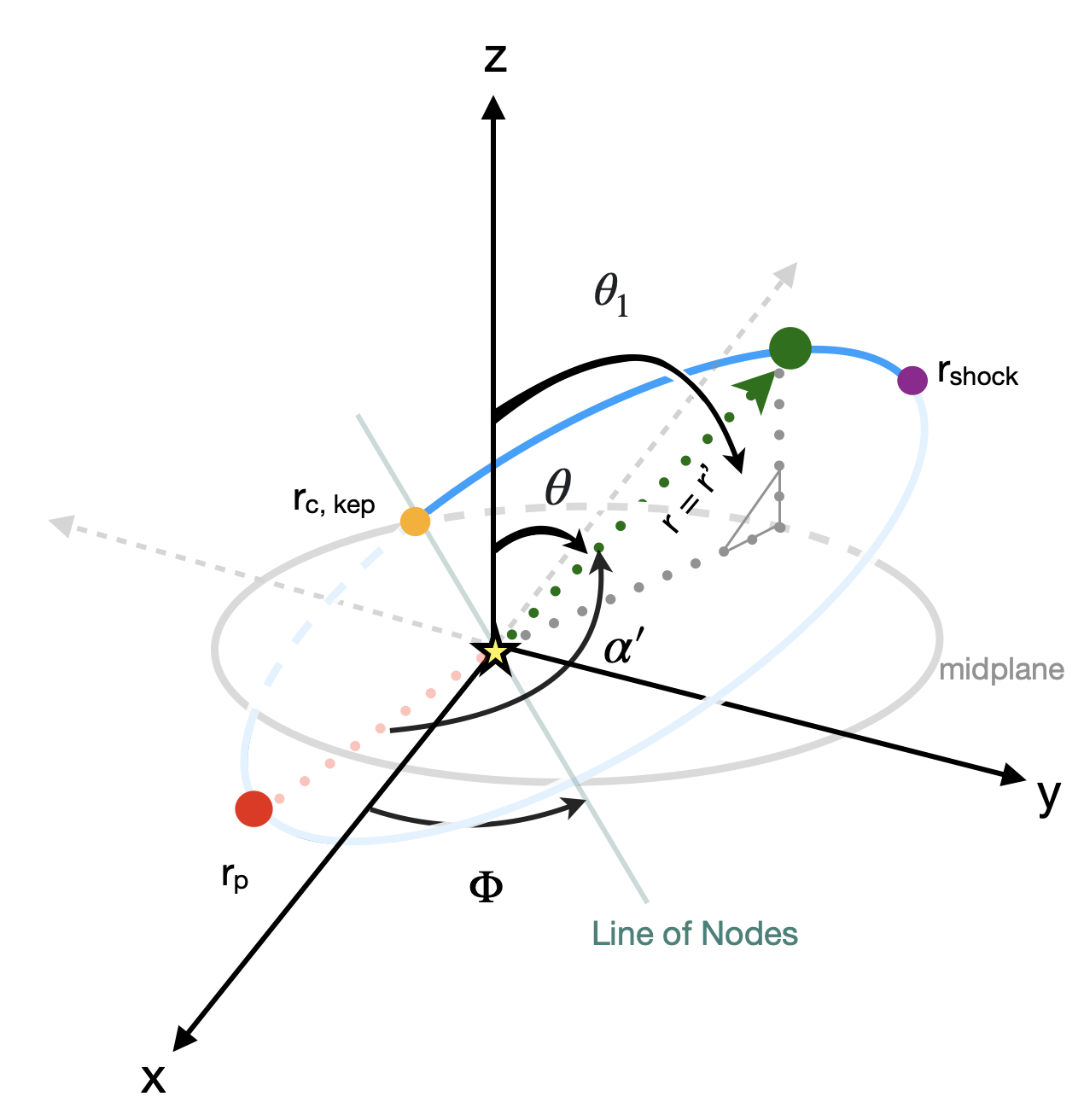}
\caption{Schematic diagram of an elliptical streamline orbit such that a gas parcel would experience a shock at the disk-envelope boundary denoted by $r_\mathrm{shock}$ (purple dot) and flow (solid blue line) towards the disk midplane at $r_\mathrm{c,kep}$ (gold dot). This diagram is described both in plane polar coordinates ($r'$, $\alpha'$) and spherical coordinates ($r$, $\theta$, $\phi$). Angle definitions are specified in Table \ref{tab:parameters_description}.}
\label{fig:ellipticalstreamline}
\end{figure}

\subsection{Elliptical gas streamlines within the disk}
\label{sec:elliptical_streamlines}

The geometry of an elliptical streamline is shown in Figure \ref{fig:ellipticalstreamline} and closely follows that shown in Figure \ref{fig:parabolicstreamline} for the parabola describing a free-fall ballistic trajectory. An infalling gas parcel transitions from a parabolic to elliptical orbit at the shock location $r_\mathrm{shock}$ that defines the disk surface, and is assumed coincident with the apoapsis point of the ellipse shown in Figure \ref{fig:ellipticalstreamline}. Polar angle $\theta_1$ specifies the orbital plane of the ellipse with respect to the z-axis. 
An individual shock location is therefore a position that connects a free-fall ballistic trajectory, or parabolic streamline orbit, and an elliptical orbit, to make a continous streamline. Applying the updated ram pressure boundary condition (see equation \ref{eq1}) returns the disk-envelope boundary shock locations in spherical coordinates and are hereafter referred to as $(r_\mathrm{shock}, \theta_1)$. The azimuthal angle coordinate $\phi_1$ at the shock is obtained using equation \ref{eq3} for the relevant branch of the parabola.

Elliptical streamlines are generated using these shock locations which delineate respective apoapsis points. 
Since elliptical streamlines transition from the free-fall ballistic trajectories, the parameter $\theta_0$ is needed in order to generate an elliptical streamline.
In this work, we determine $\theta_0$ by first using the ram pressure boundary condition to generate ($r_\mathrm{shock},\theta_1$), and then using an iterative scheme to find $\theta_0$ using equation (\ref{eq4}) for the corresponding parabolic streamline. 

In plane polar coordinates an arbitrary point ($r'$, $\alpha'$) along an elliptical streamline follows the usual definition of an ellipse where $\alpha'=0$ corresponds to periapsis, 
\begin{equation}\label{eq_ppelliptical}
r'=\frac{\ell_e}{1+e\cos(\alpha')}, 
\end{equation}
and where $\ell_e$ is the semi-latus rectum of the ellipse (line of nodes), occurring at $\alpha' = \pm \pi/2$ where the orbit crosses the midplane.  

For a thin disk $\sin(\theta_{1})\approx 1$ for the orbital plane, so that the angular momentum normal to elliptical orbit, $\Gamma_{n,e}$, is nearly equal to the cylindrical angular momentum, via $\Gamma=\Gamma_{n,e} \sin(\theta_{1}) \approx \Gamma_{n,e}$ thus leading to $\ell_e \approx r_\mathrm{c,kep}$ where $r_\mathrm{c,kep}$ is the centrifugal radius for each streamline. Full details are given in Appendix \ref{app:ang_mom}, which moreover derives the equations defining the elliptical orbit. 
Calculating the centrifugal radius for each streamline, located in the disk midplane with $\alpha'=\pi/2$ leads to,
\begin{equation}\label{eq8}
r_\mathrm{c,kep}=r_{d} \sin^{4}(\theta_{0})=a(1-e^{2}),  
\end{equation}
along with the relations defining the ellipse,
\begin{eqnarray}
(1-e) = \frac{r_\mathrm{c,kep}}{r_\mathrm{shock}}, \label{eqn_e}\\
a = \frac{r_\mathrm{shock}}{(1+e)}, \label{eqn_rshock}
\end{eqnarray}
where $a$ is the semi-major axis and $e$ the eccentricity. The term $\sin^{4}(\theta_{0})$ follows from the earlier definition of $f^{2}(\theta_{0})$ provided in \citet{Cassen_1981} and is further discussed in Appendix \ref{app:ang_mom}.

To render the position coordinates along an elliptical orbit, we first transform Equation \ref{eq_ppelliptical} to Cartesian coordinates $\vec{x}'$, having components $(x',y',z')=(r'\cos({\alpha'}),r'\sin({\alpha'}),0)$. 
We then perform a transformation using the Euler angle rotation matrix $\widetilde{A}$, such that $\vec{x}=\widetilde{A}\vec{x}'$ (using Equation 4.47 of \citet{Goldstein2002}). For our modeling, we set Euler angle $\Psi=-\pi/2$ which rotates coordinates in the orbital plane so that they match the alignment of the parabola that was adopted in the previous section.  Furthermore, since the orbital rotation axis is defined as perpendicular to the orbital plane (defined by $\theta_{1}$), then Euler angle $\Theta=90^{\circ}-\theta_{1}$.
To illustrate different azimuthally symmetric streamlines, one may use the Euler angle $\Phi$ as included in the equation.
A point on an elliptical streamline orbit can then be described by the position coordinates, 

\begin{equation}\label{eq_poscoord}
 \left( \begin{array}{c}
x \\
y  \\
z
\end{array} \right) 
=
 \left( \begin{array}{c}
~r'\cos(\alpha') \cos(\Theta) \sin(\Phi) + r'\sin(\alpha') \cos(\Phi) \\
-r'\cos(\alpha') \cos(\Theta) \cos(\Phi) + r'\sin(\alpha') \sin(\Phi) \\
- r'\cos(\alpha') \sin(\Theta)
\end{array} \right) 
\end{equation}

For the Kepler problem, the velocity in the elliptical orbit plane can be expressed in closed form, relative to the speed $v_{0}$ of a gas parcel in a circular orbit with the equivalent angular momentum $\Gamma_{n,e}$ and semi-major axis $a$. For a thin disk, namely having $sin(\theta_{1})\approx 1$,   
\begin{equation}\label{eq13} 
v_{0}=\frac{r_\mathrm{shock}v_{\phi}(r_\mathrm{shock})}{a}=(1+e) \bigg(\frac{GM}{r_\mathrm{shock}} \bigg)^{1/2} \sin(\theta_0),
\end{equation}
where we have made use of both equation \ref{eqn_rshock} and  evaluated equation \ref{eq7} for $v_{\phi}$ using $\theta = \theta_1$.

The velocity components along the plane polar coordinate system of an elliptical streamline (see equation \ref{app_eqn_vcomponents}) are therefore described by,

\begin{equation}\label{eq14}
v_{r'} = \frac{e v_{0} \sin(\alpha') }{(1-e^{2})} \qquad v_{\alpha'}=\frac{a v_{0}}{r'}.
\end{equation}

Following equivalent transformation procedures previously described, the Cartesian velocity components along an elliptical streamline are 
\begin{equation}
\begin{aligned}\label{eq15}
\begin{pmatrix}
v_x \\
v_y  \\
v_z
\end{pmatrix}
&=
\begin{pmatrix}
v_{x'} \sin(\Phi) \cos(\Theta) + v_{y'} \cos(\Phi)  \\
-v_{x'} \cos(\Phi) \cos(\Theta) +v_{y'} \sin(\Phi) \\
-v_{x'} \sin(\Theta)
\end{pmatrix} \\
\mathrm{where, } \\
v_{x'} &= v_{r'} \cos(\alpha') - v_{\alpha'} \sin(\alpha'), \\
v_{y'} &= v_{r'} \sin(\alpha') + v_{\alpha'} \cos(\alpha'). 
\end{aligned}
\end{equation}

In spherical coordinates, the velocity components are shown below.  
\begin{equation}\label{vmatrix}
 \left( \begin{array}{c}
v_r \\
v_\theta  \\
v_\phi
\end{array} \right) 
=
 \left( \begin{array}{c}
\sin(\theta) (v_x \cos(\phi) + v_y \sin(\phi)) + v_z \cos(\theta) \\
\cos(\theta) (v_x \cos(\phi) + v_y \sin(\phi)) - v_z \sin(\theta) \\
- v_x \sin(\phi) + v_y \cos(\phi)
\end{array} \right) 
\end{equation}

A remaining task is to ensure that there is a continuous transition between orbits at the known $(r_\mathrm{shock}, \theta_1, \phi_1)$ shock location. This demands that the $r_\mathrm{shock}$ point along both parabolic and elliptical streamlines align with the same $\phi_1$ value, which is not automatic. The task is aided by recognizing that for a thin disk, the line of nodes for the ellipse is nearly at right angles to the line of nodes for the parabola, so that apoapsis for the ellipse will occur near the disk midplane where the parabola crosses. Continuity is accomplished by generalizing 
the use of azimuthal shock angle $\phi_1$ in order to determine $\phi_{\mathrm{of}}$, an offset angle.

As an example, Figure \ref{fig:effectivepotential} shows our fiducial case. Using equation \ref{eq3} with parameters from Table \ref{tab:fiducial_parameters}, leads to $\phi_1=167.7^{\circ}$ at the shock location and for the correct branch of the parabola. This differs from the adopted $90^{\circ}$ alignment, thus leading to $\phi_{\mathrm{of}}= \phi_1 -90^{\circ} = 77.7^{\circ}$ for the offset angle. 
In the fiducial case, we introduce $\Phi = \phi_{\mathrm{of}} = 77.7^{\circ}$ in equation \ref{eq_poscoord} to connect the orbits by rotating the ellipse.
This places the $\bf{r_\mathrm{shock}}$ point at (the now matching) azimuthal coordinate angle $\phi=167.7^{\circ}$.

The described procedure provides a way to connect the parabolic and elliptical orbits for a given streamline. However, in practice, the calculation of $\phi_1$ was not needed for the results we present (except Figure \ref{fig:effectivepotential}), because the assumed source geometry is azimuthally symmetric.

\begin{deluxetable}{cccccc}
\tablecaption{Fiducial streamline parameters.}
\label{tab:fiducial_parameters}
\tablewidth{0pt}
\tablehead{
\colhead{$e$} & \colhead{$a$} & \colhead{$r_\mathrm{shock}$} & \colhead{$r_\mathrm{c,kep}$} &  \colhead{$\theta_0$} & \colhead{$\theta_1$}  
}
\startdata
0.43 & 52.10 au & 74.53 au & 42.44 au & 60.15$^{\circ}$ & 83.03$^{\circ}$  \\
\enddata
\end{deluxetable}

\section{Modifications to RadChemT}\label{sec:modifications_radchemt}

To generate realistic simulations we employ the RadChemT package based on the modeling method first presented by \citet{Flores-Rivera_2021}. We then discuss our updates to a standalone code within the RadChemT package which includes an updated ram pressure boundary condition that defines the disk edge, and velocity reassignments to the protostellar disk based on our modeled elliptical trajectories of gas flow from the shock at the disk-envelope boundary to the disk midplane. Aside from these updates, we follow the same prescriptions given in \citet{Flores-Rivera_2021} and that are based on the sole use of the TSC collapse model. 

RadChemT combines the gas dynamics, radiative transfer and gas-grain chemical abundance calculations to generate predictions that can be directly compared with observations. We adopted the same physical and chemical structure method as in \citet{Flores-Rivera_2021} to now include the updated ram pressure boundary condition. The TSC collapse model with an outflow cavity provides the velocity and density profile of the source that is used in \textsc{HOCHUNK3D} \citep{Whitney_2013} to obtain the dust and gas temperature profiles \citep[see Table 1 and Section 2 in ][ for the adopted physical prescriptions and parameters]{Flores-Rivera_2021}. Our disk density structure in hydrostatic equilibrium is defined with a radial power law and a vertical Gaussian structure. Then, the density and temperature structures of both the envelope and disk serve as input parameters to calculate the time-dependent chemical abundances for the gas tracer of interest. In this paper, we adopt the same C$^{18}$O abundance file to generate the synthetic moment maps used to compare our previous and current models, depicted in Figures 6-10 as circular and elliptical motions, respectively. 

The computational requirements to generate a STAK model from the equivalent RadChemT circular-orbit disk model were modest after the code development. From an operational perspective, most of the time for a RadChemT model was spent in the initial parameter fitting using the Monte Carlo radiative transfer (MCRT) code, to calculate the dust temperature from the assumed density structure and protostar luminosity. This step employs standard techniques \citep{Whitney_2013} to fit the observed spectral energy distribution. Then, velocity information is specified using the TSC collapse model within the envelope, and assuming circular Keplerian orbits within the disk. Chemical abundances are calculated using an astrochemistry code (a step that requires significant computation time), followed by a visualization step using a spectral line radiative transfer code. To create a STAK model from the original RadChemT model we 1) define a new disk surface consistent with the shock using equation (\ref{eq1}), then 2) run the MCRT code a second time to update density and temperature near the new disk surface, and 3) specify new velocities within the disk based on elliptical orbits. The STAK model steps 1-3 are based on the equations described within this work, and are largely automatic once implemented.

\begin{figure*}[htp!]
\centering
\includegraphics[width=9cm]{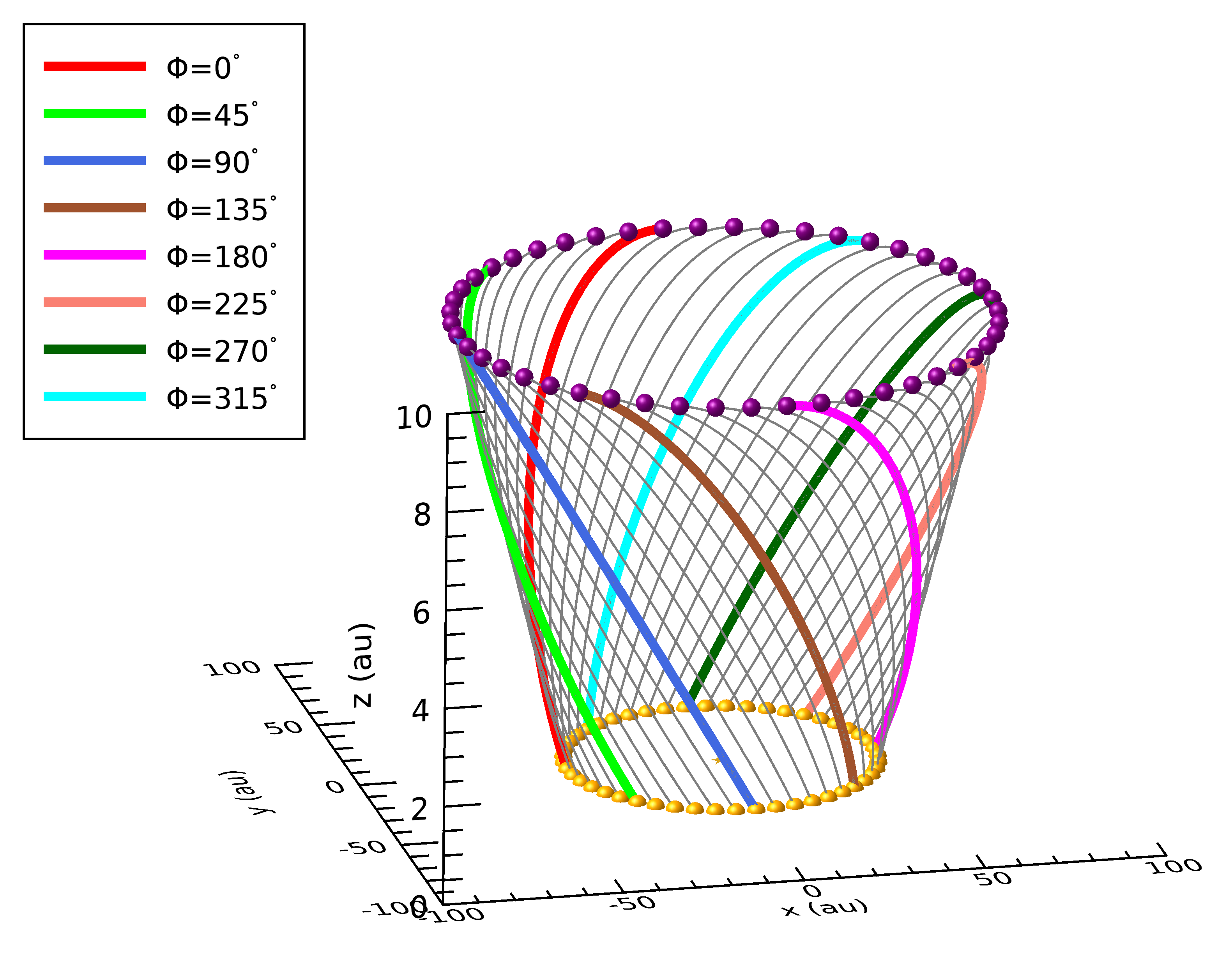}
\includegraphics[width=8cm]{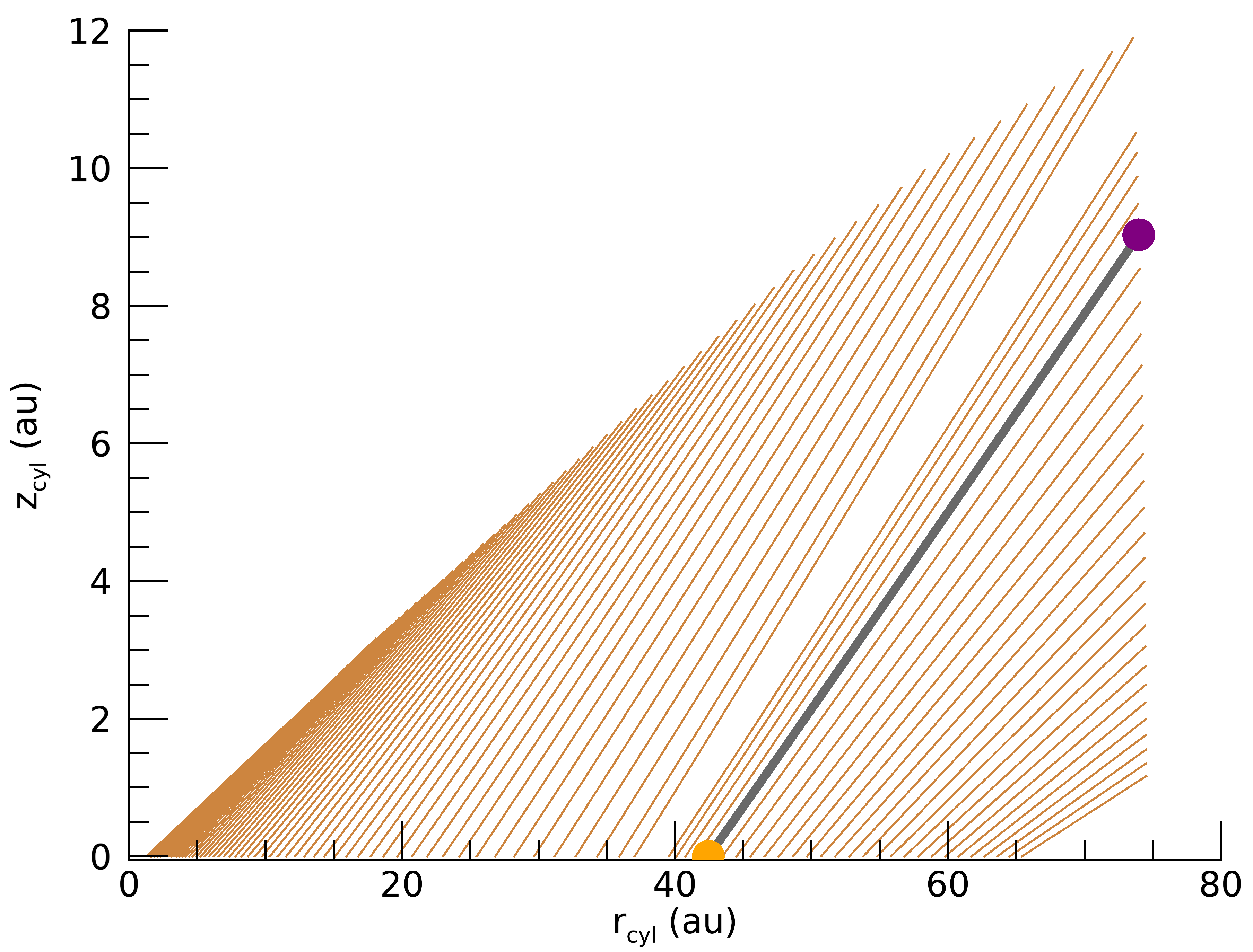}
\caption{\textit{Left} A single elliptical streamline in the disk, repeated with many azimuthal offsets in $\Phi$. \textit{Right} Various meridional streamlines for a 75 au disk radius. In gray is the equivalent streamline shown in the \textit{left} figure, but as a 2D meridional projection. In both plots, the purple point represents the shock at the disk-envelope boundary ($r_\mathrm{shock}$) and the gold point represents the shock in the disk midplane $r_\mathrm{c,kep}$. Only the top half of the disk is shown.
\label{fig:basket_and_meridional}}
\end{figure*}

\subsection{Disk Surface From Updated Ram Pressure Boundary Condition} \label{sec:disk surface}

The updated disk-envelope boundary is modified in \textsc{HOCHUNK3D} to introduce a shock prescription to the RadChemT code such that infalling gas meets a ``brick wall" at the disk surface. Before, a ram pressure boundary condition was defined considering only gas flow moving perpendicular to the disk midplane $v_{\perp}$, \citep[see equation 1 from][]{Flores-Rivera_2021}. In that work, the ram pressure in the envelope was equated to the gas pressure in the disk. This implementation led to a disk with a fairly constant opening angle out to the $r_d$ disk radius. However, this formulation for the ram pressure boundary condition had the disadvantage that $v_{\perp} \rightarrow  0$ for the important $\theta_{0}=90^\circ$ streamline where infalling gas approaches the disk from within the midplane. 
In the current work, we update the ram pressure boundary condition to include the inward velocity component $v_{env}$, but not the azimuthal $v_{\phi}$ component. The form is similar to that used by \citet{Aso_2017} to evaluate shocks at the disk edge, but has the flexibility to handle off-midplane streamlines. The updated ram pressure boundary condition is described by the equation shown below, 
\begin{equation}\label{eq1}
\rho_{disk} c_{s_{disk}}^2 = \rho_{env}( c_{s_{env}}^2 + v_{env}^2)
\end{equation}
The thermal sound speed is defined by $c_{s}=\big(\frac{P}{\rho}\big)^{1/2}=\big(\frac{kT}{\mu m_{H}}\big)^{1/2}$, the squared velocity in the envelope is described by $v_{env}^{2}= v_{r_{env}}^{2} + v_{\theta_{env}}^{2}$, and $\rho$ is the mass density. To implement Equation \ref{eq1} to determine the disk surface, the grid of temperature and density values in RadChemT are used evaluate the terms point-by-point to find the set of $(r_\mathrm{shock},\theta_1)$ locations that taken together best satisfy the equation and thus define the disk surface. A particular feature of \textsc{HOCHUNK3D} made implementation simple. Namely, \textsc{HOCHUNK3D} calculates both $\rho_{disk}$ and $\rho_{env}$ as well as temperature for each grid point in the relevant region. Thus, the ram pressure terms in equation \ref{eq1} can be evaluated so that a grid point could be assigned as belonging to either the disk or envelope. The disk surface (i.e. $r_\mathrm{shock}$), where ram pressure equality holds in equation \ref{eq1}, was then defined by interpolating between the two regions.

Within the disk, each $(r_\mathrm{shock},\theta_1)$ pair occurs at apoapsis for motion following an elliptical orbit. For the infalling envelope, setting $r = r_\mathrm{shock}$ and $\theta = \theta_{1}$ in Equation \ref{eq6} gives $\theta_{0}$ for the corresponding parabolic orbit that terminates at $r_\mathrm{shock}$. Model parameters M and $r_\mathrm{d}$ (see Eqn \ref{eq5}) are also needed to define a streamline. Table \ref{tab:fiducial_parameters} illustrates the result for the fiducial streamline.

This new prescription (Eqn \ref{eq1}) using inward $v_{env}$ at the disk surface naturally leads to the $v_{\phi}$ component being constant across the shock, thus conserving angular momentum. 
While orbits momentarily appear circular at the shock, the value of angular momentum corresponding to $v_{\phi}$ is not correct for a circular Keplerian orbit.
Instead, as we argue in \S \ref{sec:dynamicdisk}, the gas accelerates inward due to gravity, thereby increasing the magnitude of the radial and polar velocity components, and transitioning to the elliptical orbit described in \S \ref{sec:elliptical_streamlines}. The velocity of the elliptical orbits leads to a very different pattern of Doppler velocity than is seen for circular orbits. In \S \ref{sec:results} following, we present the new Doppler velocity pattern for the dynamical STAK disk.

We treat the gas flow as transitioning immediately to elliptical streamlines on crossing the shock.  This is valid if after the abrupt heating at shock passage, the gas cools rapidly enough that pressure gradients cannot significantly deflect the flow.  Rapid cooling is consistent with results presented in \citet{Neufeld_1994}, adjusting for the shock speeds below 10 \kms\  prevalent in the outer disk.  To further test this assumption, we compute the immediate post-shock temperature and pressure using the jump conditions, and the subsequent cooling timescale using thermal emission from the dust component, as set out in Appendix~\ref{subsec:coolingt}.  The result is that for streamlines reaching the disk's outer half, the gas temperature peaks below a few hundred Kelvin and returns to near the pre-shock level within an hour.  The flow is thus transonic for a tiny fraction of an orbit and supersonic everywhere else, with gravity the main force and thus the orbits well-described by conic sections.

\section{Results} \label{sec:results}

\begin{figure*}[htp!]
\centering
\includegraphics[width=18cm]{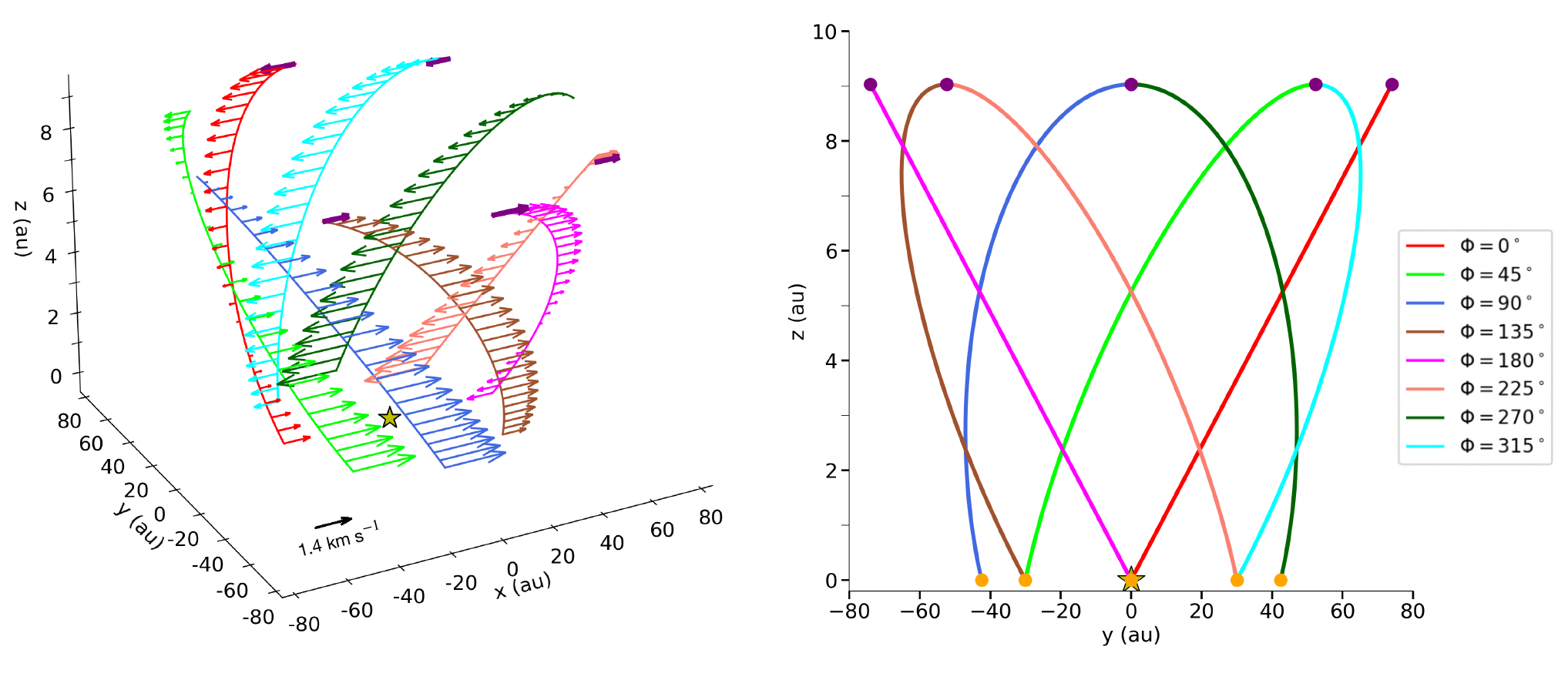} 
\includegraphics[width=8cm]{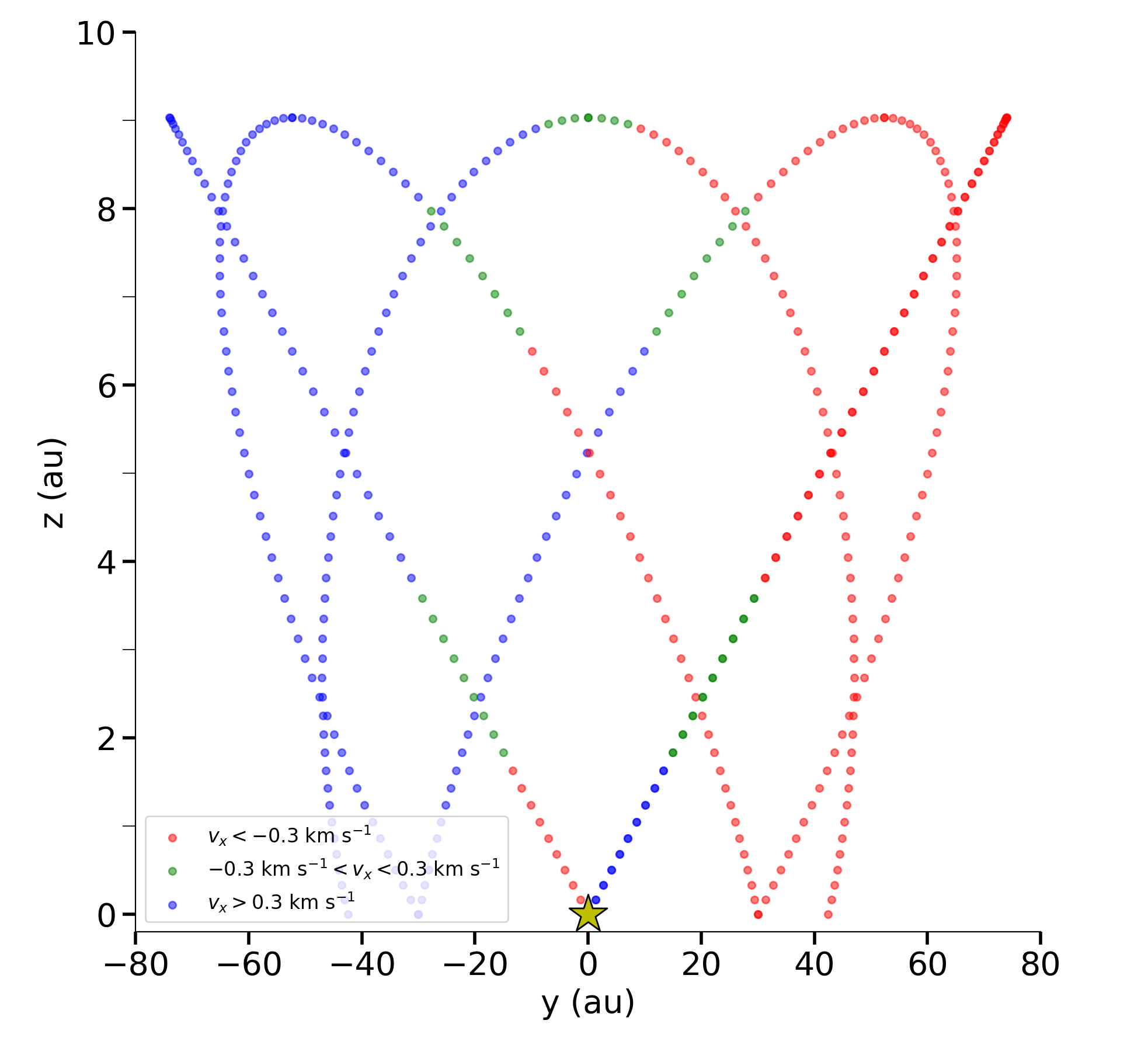}
\caption{Velocity structure corresponding to single streamline case with 7 offsets in $\Phi$ (Fig. \ref{fig:basket_and_meridional}). \textit{Top left} the velocity vectors represent the $v_x$ component, seen in a perspective view. \textit{Top right}, shows the projected streamlines as viewed from the positive x-axis. \textit{Bottom}, shows the corresponding Doppler velocity projection, as viewed from the positive x-axis. Markers show red and blue shifted gas along with gas containing speeds in an intermediate regime (green) of $\pm$0.3 km s$^{-1}$. The star symbol locates the protostar. }
\label{fig:streamlines_offsets}
\end{figure*}

\subsection{Elliptical Streamlines: Symmetry and Velocity}

Now we extend results to three dimensions, first limiting discussion to the disk as it is the only region containing new velocity structure. To visualize and exploit the azimuthal symmetry of elliptical streamlines, Figure \ref{fig:basket_and_meridional} (\textit{left}) shows elliptical streamlines in the disk seen as a basket-like shape once the transformation matrix $\widetilde{A}$ for rotation through the Euler angle $\Phi$ is applied to the fiducial streamline case (see \S \ref{sec:elliptical_streamlines} and Eqn \ref{eq_poscoord} discussion). The basket is cone shaped because the underlying ellipses are conic sections. The purple points in the figure represent respective $r_\mathrm{shock}$ points while the gold points represent $r_\mathrm{c,kep}$ where gas parcels settle into the disk midplane. Note that the disk is indeed thin; for better clarity of vertical structure the z-scale has been expanded. Also notice that azimuthal symmetry means that none of the streamlines cross before the gas parcels reach the disk midplane. 

For the fiducial case having $\theta_{0}=60^\circ$, the polar angle $\theta_{0}$ defines the orbital plane of the parabolic part of the streamline (not shown), transitioning to an elliptical orbit at the shock location (purple dot), which occurs at $(r_\mathrm{shock},\theta_1)$ in spherical coordinates. The shock location also corresponds to apoapsis for the elliptical orbit; because of this then $\theta_1$ also gives the orbital plane of the ellipse, because the orbit plane is then defined using the apoapsis location at the shock $(r_\mathrm{shock}=74~\mathrm{au}, \theta_1=83^\circ$; Tbl. \ref{tab:fiducial_parameters}) and the protostar position at the origin.  

Generalizing from Figure \ref{fig:basket_and_meridional} (\textit{left}) that shows the fiducial case; other streamlines corresponding to different $\theta_1$ polar angles will look similar, appearing as a nested set of baskets, located both interior and exterior. To further compare elliptical streamlines having different $\theta_1$ polar angles, Figure \ref{fig:basket_and_meridional} (\textit{right}) employs a meridional projection, useful for axisymmetry, that is a 2D projection into cylindrical coordinates (r$_\mathrm{cyl}$, z$_\mathrm{cyl}$, any $\phi$). The disk surface is visible in outline along the upper and right boundary. Each straight line is a streamline having a different $\theta_1$ polar angle, with the top of the line at the shock location, and the bottom of the line at the disk midplane. There is an apparent gap landing at $\sim 37~au$ arising from the way we implemented the ram pressure boundary condition. Namely we located the top and outer edge of the disk separately based on a cylindrical coordinate grid search. However, the top right corner of the disk was pointlike and hard to locate. A future modification to RadChemT will implement a radial based grid search to mitigate the numerical effect. 

The \textit{right panel} showing our fiducial elliptical streamline case is color-coded in grey, with the grey straight line (right panel) corresponding to the straight side of the basket shown in the left panel. Each streamline in the model has its own angular momentum that is constant throughout the elliptical orbit. The meridional projection (right panel) demonstrates that the streamlines do not cross. Formally, streamlines do not cross because $r_\mathrm{c,kep}$ increases monotonically with parabolic $\theta_0$ polar angle. This is due to angular momentum increasing outwards in the adopted TSC cloud core initial state.

We briefly discuss several nonphysical situations where intersecting or orphan streamlines occurred. Initially, the disk surface was defined using the spatial grid point locations. However, the numerical coarseness of the grid led to roughness in the disk surface, which in turn led to crossing streamlines. This behavior was successfully suppressed by interpolation of the $r_\mathrm{shock}$ locations. One exception is near the disk edge, where the density is known to have a singularity \citep{Cassen_1981}. To compensate, the short elliptical streamlines with intersection and having $\theta_1\geq89^{\circ}$ were omitted in our study. Small $\theta_0$ polar angles were also problematic, due to the assumed shape of the outflow cavity and jet. The smallest $\theta_0$ that avoids the outflow and thus produces a continuous streamline that meets the disk occurs at $\theta_0=21^{\circ}$. This corresponds to a disk region at approximately 11 au distance from the protostar. Inside this disk region, material infalling from the envelope cannot directly intersect the disk. We performed tests using several different assumptions for the velocity behavior in the special regions, and found that the results we present are not affected. 

So far, we have explored multiple possible trajectories for a single elliptical streamline both in 2D and 3D under azimuthal symmetry after an initial shock on the disk surface. Now we explore the velocity components under this symmetry, with the aim of displaying the Doppler velocity. Equation \ref{vmatrix} describes the velocity components of the elliptical streamlines.  Figure \ref{fig:streamlines_offsets} shows the same color-coded elliptical streamlines as in Figure \ref{fig:basket_and_meridional} for consistency. The direction of the space velocity is tangent to the curve at each point, for each streamline. However, the Doppler effect is only sensitive to part of the velocity, namely the velocity component that is projected along the line of sight to the observer. To illustrate the Doppler effect, the purple arrow shown perpendicular to the top of each individual elliptical orbit in the 3D plot (Fig. \ref{fig:streamlines_offsets} \textit{top left}) shows velocity corresponding to the initial shock location ($r_\mathrm{shock}$). Then, as the gas parcel travels approximately one-quarter of the full elliptical orbit, it reaches the second shock location at the midplane ($r_\mathrm{c,kep}$). The perpendicular arrows with respect to every individual color-coded orbit show the Cartesian x-component of the velocity vectors as described in Equation \ref{eq15}. This represents the Doppler velocity that an observer located on the positive x-axis would detect. 

The 2D plot (Fig. \ref{fig:streamlines_offsets} \textit{top right}) shows the same streamlines as in the \textit{left} plot, but projected in the z-y plane. Purple and gold markers delineate respective $r_\mathrm{shock}$ and $r_\mathrm{c,kep}$ locations. The elliptical streamlines as seen in the \textit{right} plot are the observer's view, projected in the plane of the sky, if the disk were in an edge-on configuration. The velocity vectors and the geometrical configuration of the streamlines give a preview of the Doppler motions that go into creating a synthetic 2D image containing the spatial information of the gas distribution in the disk. 

To further show the relative motion of the gas towards the line-of-sight, we produce the Doppler velocity projection in the \textit{bottom} of Figure \ref{fig:streamlines_offsets}. The velocity streamlines and viewing angles are the same as in Figure \ref{fig:streamlines_offsets} (\textit{right}) plot. The dot markers color-coded in blue represent the gas moving towards the observer's line of sight, whereas those in red represent the gas moving away from us. The green dots show the regions where the gas parcels move at intermediate speeds of $\pm$0.3 km s$^{-1}$, close to the velocity of the protostar.  Considering the case of a disk with circular orbits, the observed Doppler pattern is expected to show red-shifted velocity on the right, blue-shifted on the left, and a bar of green near the system velocity in a vertical band near the rotation axis. In contrast, elliptical orbits lead to unexpected velocities that cross the rotation axis to appear in the ``forbidden zone'', of some blue-shifted gas appearing in the red-shifted quadrant (red and light-green streamlines), and vice versa for some red-shifted gas appearing in the blue-shifted quadrant (magenta and orange streamlines). The visualization in Figure \ref{fig:streamlines_offsets} displays an edge-on disk inclination ($i = 90\degr$); other source inclinations follow the same process, although the detailed features of the velocity projection may differ. 

\begin{figure*}[htp!]
\centering
\includegraphics[width=18cm]{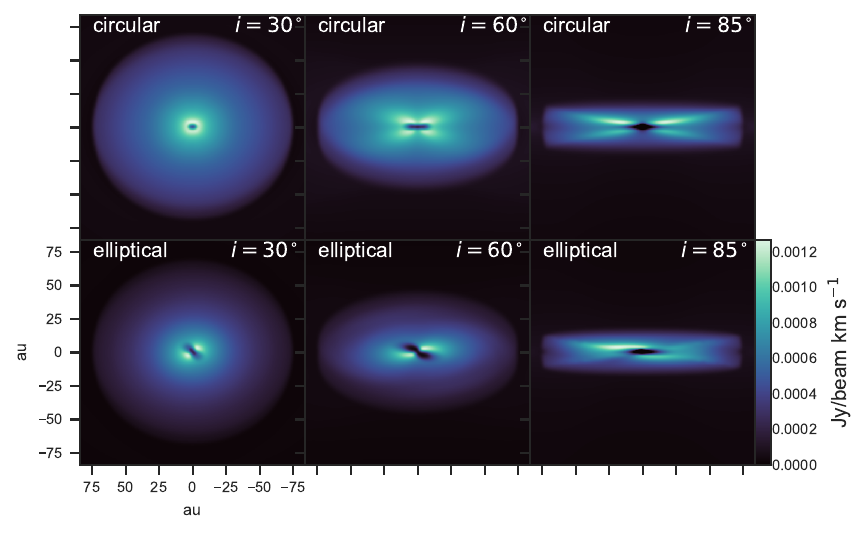} 
\caption{Integrated intensity (Moment 0) maps of \cuo{18}\ the fiducial case for the standard circular orbit disk versus the elliptical orbit disk from our models. Three source inclinations are compared. The images are continuum subtracted. The protostar and disk parameters used are $M_{*}$=0.22 $M_{\odot}$ and $r_{d}$=75 au. The image spatial resolution is 2.5~au.}
\label{fig:moment0_maps}
\end{figure*}

\subsection{Constructing Synthetic ALMA Spectral Line Cubes} \label{sec:constructing cubes}
To compare the STAK dynamic disk predictions with high spatial resolution millimeter interferometry data, such as from ALMA, we construct synthetic spectral line cubes and moment maps using RadChemT (see \S \ref{sec:modifications_radchemt}). The reference model based on \citet{Flores-Rivera_2021} contains both continuum and C$^{18}$O(2-1) spectral line emission at 219 GHz, an adopted source radial velocity of 6.0 \kms, with 1~au pixels and 100 frequency channels having 0.168 \kms velocity resolution. 

Following standard practice, the continuum emission was subtracted from the model cube in order to produce a spectral line model cube. The continuum image (not shown) was constructed from off-line channels by averaging the first and last channel of the spectral line cube, meant to contain no spectral line emission and therefore show only thermal dust emission. The C$^{18}$O spectral line cube was constructed by subtracting the continuum image from each plane of the cube, thus showing just C$^{18}$O gas when the emission is optically thin. In practice, the emission in the near vicinity of the protostar is often not optically thin, potentially leading to absorption or negative features in the center of the image after continuum subtraction. Sometimes the effect is mitigated in moment maps by excluding velocity channels, typically channels near the systemic velocity. However, in this work, no velocity channels have been excluded in the moment maps presented. Further information is available in the case of the protostar L1527, where \citet{vanthoff_2018} provide an analysis of optical depth versus Doppler velocity. 

Moment 0 ($\int I_{\nu}d\nu$) integrated intensity and moment 1 velocity ($\int v I_{\nu}d\nu$) images  for C$^{18}$O were then constructed within $\pm 4~\kms$ of the $6~\kms$ source velocity, namely by integrating spectral line emission over a velocity range of [2,10] \kms. The full equations for the velocity moments, involving sums and normalization factors, are available on the ALMA website \citep{McMullin_2007}. The equation for the classic moment 1 velocity is sensitive to the spectral line shape, and is implemented here. This is in contrast to moment 9 or quadratic moment methods that prioritize measuring the spectral line centroid \citep{McMullin_2007, Teague_2018a}, and which may have a different behavior for the expected double-peaked spectral lines.

For visualization, three different source inclination angles were selected to span the range of results. The $i = 30\degr$ inclination shows a near face-on case (having lowest projected velocity), while $i = 85\degr$ presents a near edge-on disk (highest projected velocity). For randomly oriented source rotation axes, the average inclination is $i = 60\degr$; therefore the middle value $i = 60\degr$ displays the typical appearance.

\subsection{Emission Signatures in the Outer Disk}

\begin{figure*}[ht!]
\centering
\includegraphics[width=18cm]{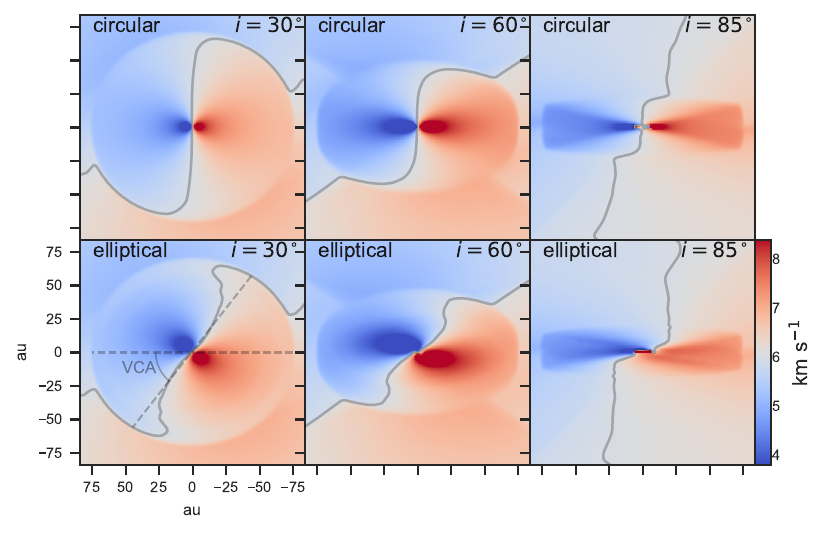} 
\caption{Averaged velocity (Moment 1) maps of the C$^{18}$O fiducial case for the standard circular disk versus the elliptical disk from our models. The images are also continuum subtracted and the disk parameters are the same as in Figure \ref{fig:moment0_maps}. The adopted system radial velocity is 6 \kms\ for the fiducial case. The image spatial resolution is 2.5~au.}
\label{fig:moment1_maps}
\end{figure*}

We present results for both the intensity moment 0 and velocity moment 1 maps; readers may skip ahead to \S \ref{sec:velocity_sig} to find the highly diagnostic and simpler velocity moment results. 
Figure \ref{fig:moment0_maps} shows moment 0 images of the integrated C$^{18}$O line emission, for disks having circular orbital motion (\textit{top row}) compared to elliptical orbital motion (\textit{bottom row}) at three different disk inclinations. Each image is continuum subtracted and has 2.5 au spatial resolution. The disk emission is emphasized, appearing in blue, while the fainter envelope emission appears black for the selected color table. The apparent disk major axis bisects the disk horizontally. From an observational perspective, the disk major axis is identified using the dust continuum emission (not shown), and is the same for both models. The apparent disk minor axis bisects the disk vertically, and from an observational perspective is often identified with the outflow axis. 

The morphology of the gas (Fig. \ref{fig:moment0_maps}) is clearly different between the top row, showing circular motion and the bottom row, showing elliptical orbital motion. For the circular case, the gas emission is near mirror symmetric with respect to the disk major axis (i.e. top/bottom symmetric), and also to the disk minor axis (i.e. left/right symmetric). By contrast, the elliptical structures give the overall impression of asymmetries, or twists, with respect to the disk major and minor axes. Further, the $i = 60\degr$ inclination has an apparent inner spiral, while $i = 85\degr$ shows strong top left/ lower right apparent brightness asymmetry, even though the modeled underlying structure in all cases is azimuthally symmetric.  

Several effects are visible in Figure \ref{fig:moment0_maps}. An important consideration is the interplay between column density and optical depth along with temperature: higher column density produces brighter emission, so sightlines with longer pathlength, e.g. higher inclination, trend brighter. However, at very high column density the gas becomes optically thick, so that the line emission becomes self-absorbed by cooler gas that is closer to the observer, found at the edge of the disk and also in the disk midplane.  This is visible in the familiar circular orbit case (top row) where the bright region becomes horizontally larger (i.e. major axis) with increasing inclination from left to right; and a self-absorption band (fainter emission) appears in the disk midplane. Note that very dark features are in many cases artifacts of the continuum subtraction step (see \S \ref{sec:constructing cubes}), due to over-subtraction by the constructed continuum image, in regions of high line+continuum optical depth. Also, because optical depth plays an important role, then using a molecule different from C$^{18}$O can affect the appearance of the moment maps. The self-absorption features described here are present in high spatial resolution observational data that fully resolve the disk; recent studies for L1527  \citep{vanthoff_2023} and IRAS 04302+2247 \citep{Lin_2023} provide examples. 


Velocity crowding of streamlines is also a consideration. The modeling method of creating a 3D space-space-velocity cube relates line-of-sight position to Doppler velocity (cube third axis). This means that longer pathlength can come about via velocity crowding in the Doppler velocity. For the circular orbit case this occurs along the minor axis extending vertically from the protostar, namely in lines-of-sight where the space velocity lies in the plane of the sky and thus the Doppler velocity goes to zero (i.e. system velocity). 


For elliptical orbit disks (Fig. \ref{fig:moment0_maps} bottom row), the change in space velocity means a different set of streamlines have zero Doppler velocity (i.e. system velocity), and appear as bright emission features for $i = 30\degr$ (bottom left) and $i = 60\degr$ (bottom middle). For $i = 30\degr$ the central bright emission is predicted to appear significantly rotated with respect to the \textbf{dust continuum} major axis. Offset in the opposite direction there is relatively little gas at high velocity, leading to a dark/faint/negative feature. For $i = 60\degr$ the central bright emission is similar but appears rotated with respect to the major axis in a loose spiral feature. The elliptical orbit models imply that observational data should display asymmetries with respect to the disk major axis, for observations that fully resolve the disk.

At nearly edge-on inclinations, additional projection effects become important for the morphology of the elliptical orbit case. Emission that arises, for example, in the top half of the disk (+z), primarily appears in the top-half of the image at $i = 85\degr$ (Fig. \ref{fig:moment0_maps} bottom right panel), thus lying above the disk major axis. This spatial offset differs from the lower inclination ($i = 30,60\degr$) cases, and leads to different emission morphology. Similarly, the bottom half of the disk primarily contributes to emission in the bottom half of the image. Asymmetry at $i = 85\degr$ (bottom right panel) for the brightest emission arising from elliptical orbits appears to the upper left of the protostar. There is similar but slightly fainter emission  to the lower right of the protostar (the color table disguises that it is only 20\% fainter). Note that at exactly $i = 90\degr$ the emission structure will be symmetric. However, for $i = 95\degr$ (not shown) the asymmetry is flipped with respect to $i = 85\degr$; \S \ref{sec:case_study} further discusses the asymmetry in the nearly edge-on case as an inclination effect. Here, the source inclination is defined using the right hand rule to establish the positive z rotation axis of the disk, which in the case of $i < 90\degr$ points towards the observer with $i = 0\degr$ for a face-on disk. Values of $i > 90\degr$ are relevant due to the asymmetric velocity induced structure.



\begin{figure*}[htp!]
\centering
\includegraphics[width=17.5cm]{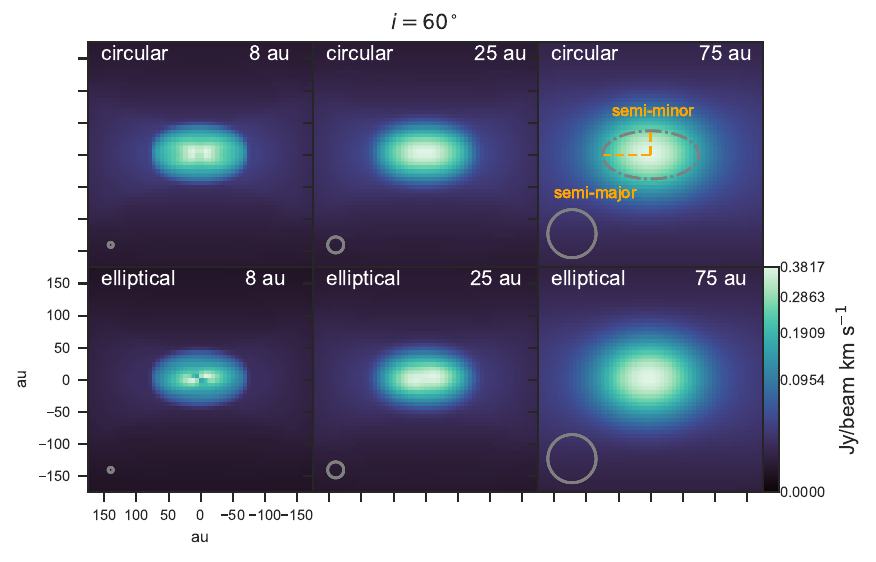} 
\includegraphics[width=17.5cm]{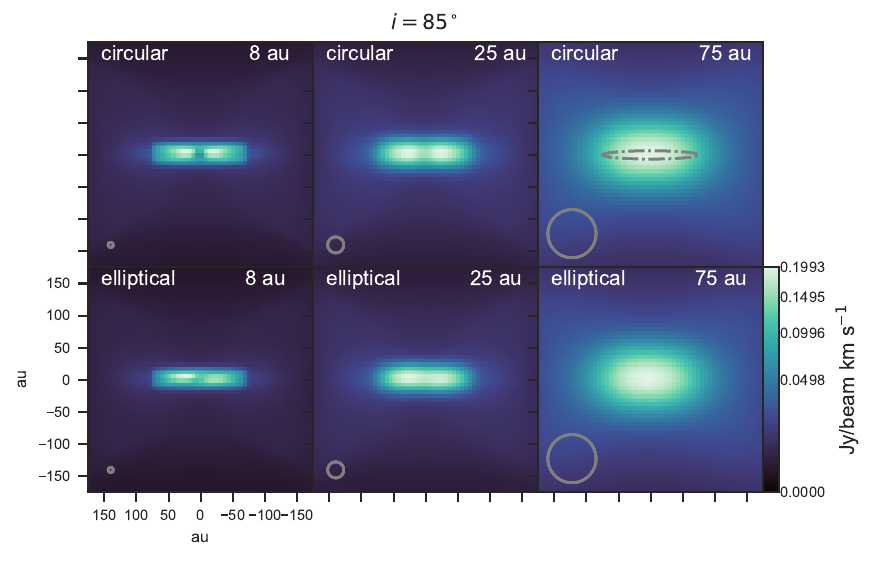} 
\caption{Same as Figure \ref{fig:moment0_maps} for $i=60^{\circ}$ (\textit{top a figures}) and $i=85^{\circ}$ (\textit{bottom b figures}). The images are convolved using a 2D Gaussian with circular FWHM = 8 au, 25 au, and 75~au at a distance of 140~pc. Overplotted in dotted-dashed grey line for both 75 au cases is the disk shape with radius 75~au for reference.} 
\label{fig:spatial_res_mom0}
\end{figure*}

\subsection{Post-shock Velocity Signatures in the Outer Disk}
\label{sec:velocity_sig}

Figure \ref{fig:moment1_maps} shows the different velocity morphologies for the circular (\textit{top}) and elliptical (\textit{bottom}) gas motion at three different disk inclinations. The grey-color solid line corresponds to 6 \kms, which is the system velocity considered in our fiducial case.  As for Figure \ref{fig:moment0_maps}, the disk major axis is at 0 au in the horizontal direction. The outline of the 75~au disk is clearly seen. Regions outside the disk boundary contain contributions only from envelope or outflow gas, while regions inside the apparent boundary contain contributions from both the disk and envelope, and potentially the outflow. 

Figure \ref{fig:moment1_maps} demonstrates that the velocity moment 1 images are highly diagnostic of differences between disks having circular orbital motions (top row) and elliptical orbital motion (bottom row), and appear less complex than the moment 0 intensity images. The system velocity (grey line) shows different morphology when comparing the two orbits for the same inclination. For the circular case, notice that the system velocity contour line is vertical, almost symmetrical, and consistent with the outflow axis. For the elliptical case, this system velocity contour line is no longer a straight line oriented along the outflow axis.  Moreover, {\it the elliptical case shows twisted kinematics in the center, resembling a spiral-like structure.} The spiral feature is most distinguishable for the elliptical case with disk inclination of $60^{\circ}$ for both moment 0 and 1 maps, and provides a possible explanation of spiral features already seen in some gaseous protoplanetary disks. Considering next the highest velocities (deepest blue or red), they no longer occur on the disk major axis (see top row), but instead on a line significantly rotated from the major axis (see bottom row). At the outer disk edge, the grey line curves sharply where it transitions to gas that is related to the outflow shell. See Appendix \ref{largeFOV} and Figure \ref{outflow_shell} for how the disk structure connects to the envelope and outflow on larger scales. 

To diagnose asymmetries associated with the STAK disk we introduce the concept of the Velocity Crowding Angle (VCA), an angle that is closely related to the bright features seen in the moment 0 images (Fig. \ref{fig:moment0_maps}) that arise near zero Doppler velocity (i.e. the system velocity) from velocity crowding effects. To provide a working definition, we define it as the angle between the disk major axis (horizontal line as defined by the continuum) and the mostly straight-line segment of the system velocity contour (solid grey line) in the velocity moment map (Fig. \ref{fig:moment1_maps}, see dashed grey lines in lower left panel). The VCA is a potential observational measure that can distinguish between the circular case, which should have VCA $=90\degr$ (apart from any outflow confusion), and the VCA $< 90\degr$ elliptical case.  For the STAK disk Figure \ref{fig:moment1_maps} shows how the VCA twist depends on inclination (bottom row, left to right). 

To detect the STAK disk, we conclude that twisted kinematics in the center with respect to disk major and minor axes should be visible in high spatial resolution observations with good signal to noise that fully resolve the disk; this Doppler effect diagnostic signature occurs even though the modeled underlying structure in all cases is azimuthally symmetric.

\begin{figure*}[htp!]
\centering
\includegraphics[width=17.5cm]{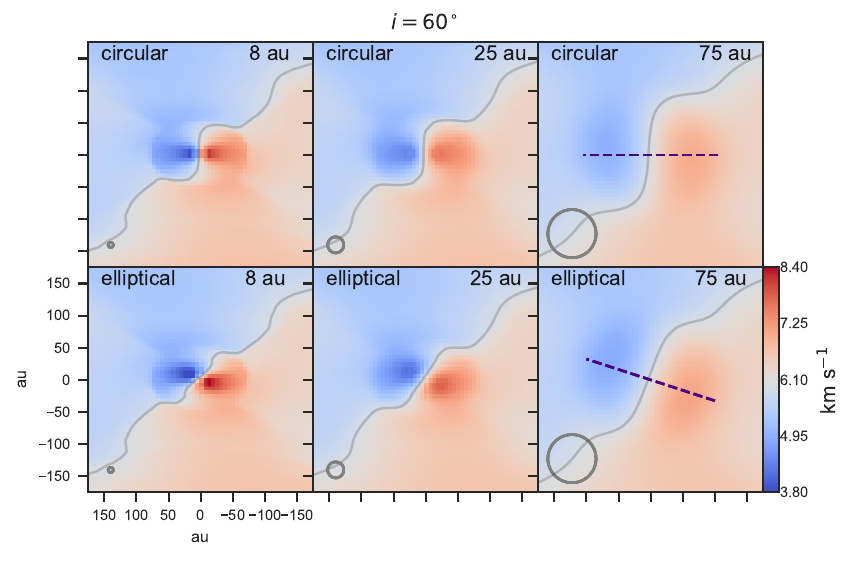} 
\includegraphics[width=17.5cm]{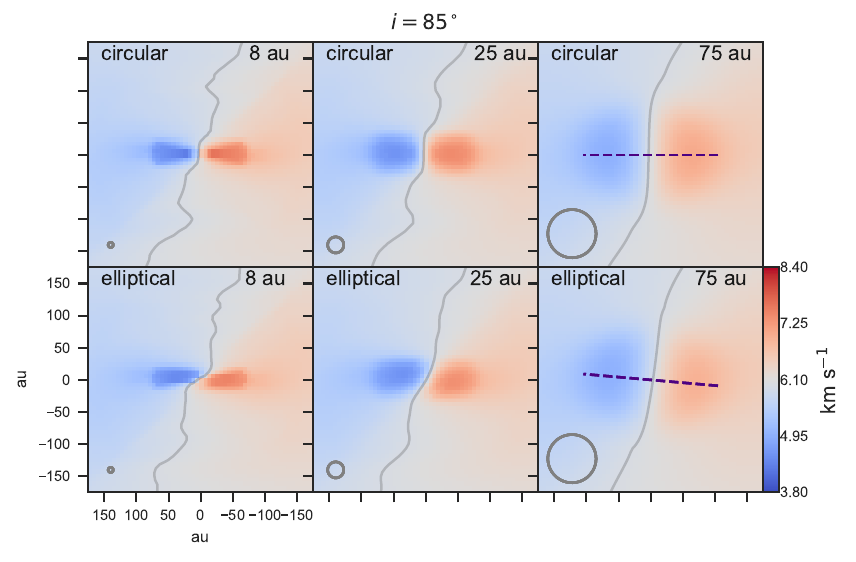} 
\caption{Same as Figure \ref{fig:spatial_res_mom0} but showing the moment 1 maps. The grey contour line corresponds to the systemic velocity (6 km~s$^{-1}$) and the dashed-purple line perpendicular to it shows the velocity symmetry for both circular and elliptical cases.} 
\label{fig:spatial_res_mom1}
\end{figure*}

\subsection{Telescope spatial resolution effect}
\label{sec:telescope_spatial_res}
 Many observational data do not fully resolve protoplanetary disks, as disks vary in size and can be quite small. Therefore this section explores the effect of telescope spatial resolution on the STAK disk model, to better illustrate its key features.
Figure \ref{fig:spatial_res_mom0} and \ref{fig:spatial_res_mom1} show the moment 0 maps and the moment 1 maps, respectively, for $i=60^{\circ}$ (\textit{top a figures}) and $i=85^{\circ}$ (\textit{bottom b figures}). These images convolve the 75~au radius disk with a Gaussian 2D function having FWHM = 8 au, 25 au, and 75 au that are meant to match ALMA beam sizes that partially resolve the disk. Adopting a circular beam is a good approximation when the the declination of the source is not high; a full exploration of beam shapes and position angles is outside the scope of this work. 

Telescope spatial strongly affects the moment 0 emission images (Fig. \ref{fig:spatial_res_mom0}). Only the well resolved disk with 8~au resolution noticeably retains spatial asymmetry in the STAK disk. Otherwise the asymmetry largely disappears, exhibiting little difference between the circular and elliptical orbit cases for 25~au and 75~au resolution. To aid in guiding the eye to the underlying source geometry, note the lines drawn in the upper right panel. The dot-dash line outlines the projected 75~au disk radius. Dashed lines show the semi-major disk axis (observationally defined by dust continuum emission) and semi-minor disk axis (assumed to be aligned with the outflow axis).   

By contrast, the moment 1 velocity images (Fig. \ref{fig:spatial_res_mom1}) appear well suited to diagnosing the velocity asymmetry that is predicted by the STAK disk. The moment 1 velocity images retain the velocity asymmetry for all three spatial resolutions and both inclinations. Diagnosing the velocity asymmetry can potentially done in two ways. First, if using the system velocity (grey solid line), then the previously discussed Velocity Crowding Angle (Fig. \ref{fig:moment1_maps} lower left panel) can measure differences between the circular case (near-vertical line) and elliptical case (line differs from vertical). 

One possible drawback is that the VCA twist depends on spatial resolution, in particular for the $i=85^{\circ}$ inclination. A second approach is to focus on the high velocity gas (dashed purple line in right panels), which no longer coincides with the disk major axis for the STAK disk. A potential drawback is that the high velocity emission is fainter and thus more affected by the noise level. Also, a caution is that different line lengths may influence the result so that, for example, the bisecting line may be definition dependent. Nevertheless, there are clearly differences between the circular and elliptical cases present in the velocity images, suggesting velocity moment images should be a promising avenue for future exploration.

\section{Discussion}

\subsection{Implications for Observations: a Case Study for L1527} \label{sec:case_study}

ALMA observations of protostars that fully resolve their disks are increasingly available. The recent eDisk survey of \citet{Ohashi_2023} presents a gallery of 1.3 mm continuum observations for dozens of protostars, where most disks show smoothly varying emission having little of the substructure that is frequently seen for older Class II disks. 

For our purpose, spectral line observations are needed. We begin by comparing our elliptical and circular modeling framework with ALMA data for the protostar L1527, since there is known evidence for noncircular motions in its outer disk. We first compare with ALMA archival data, to look for diagnostic signatures of the STAK disk that \S\ \ref{sec:telescope_spatial_res} suggests may be found in the velocity moment map, even for data that only partially resolve the disk. We then consider more recent ALMA spectral line data that have higher spatial resolution.

\begin{figure*}[htp!]
\centering
\includegraphics[width=13cm]{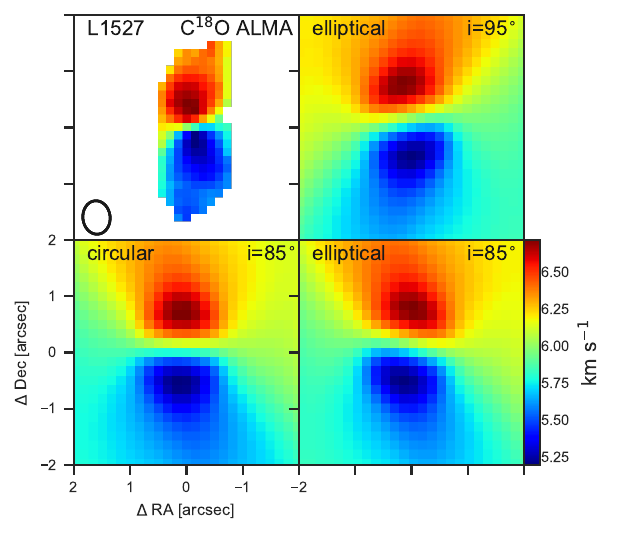} 
\caption{L1527 C$^{18}$O ALMA Moment 1 map compared with the elliptical disk model at 95$^{\circ}$ (\textit{top right}), and with the circular disk and the elliptical disk model with 85$^{\circ}$ (\textit{bottom}) at low spatial resolution. Spatial resolution in upper left panel is FWHM = 0.96$\arcsec~\times~0.73\arcsec$, while other panels show equivalent circular resolution FWHM = 0.84$\arcsec$, corresponding to 120~au. } 
\label{fig:mom1_ALMA_comparison}
\end{figure*}

We use the C$^{18}$O(2-1) ALMA data of L1527 that have a spatial resolution of 0.96$\arcsec~\times~0.73\arcsec$, enough to cover envelope scales, that was taken during cycle 0 on 2012 August 26 (Project code: 2011.0.00210.S; PI: N. Ohashi). These are compared with three models having equivalent circular resolution 0.84\arcsec, equal to the geometric mean, that derive from our previous extensive modeling and data comparison in \citet{Flores-Rivera_2021}. Figure \ref{fig:mom1_ALMA_comparison} shows the velocity moment 1 map, showing the ALMA data in the upper left panel, and the circular Keplerian disk model $i=85^{\circ}$ in the lower left panel. Notice that the disk major-axis is oriented North-South to match the data, unlike previous figures. Overall, data and model have similar morphology, demonstrating a Keplerian disk signature with red and blue shifted gas of appropriate peak velocity and spatial offset. However, the data show also asymmetries similar to the STAK disk predictions. Two model inclinations are shown for $i=95^{\circ}$ (upper right), and $i=85^{\circ}$ (lower right). Of the three models, the elliptical orbit model at $i=95^{\circ}$ (top right) seems to match most closely to the L1527 C$^{18}$O ALMA data. Both the peak red and blue shifted gas locations match, and the inner VCA twist tracing the system velocity (green) looks well aligned. The connection to the larger scale flow is strongly influenced by the bipolar outflow even in C$^{18}$O and is shown in Appendix D (Figure \ref{outflow_shell}). 

We note that velocity asymmetries arise naturally at the shock in the STAK dynamic disk, but are not a unique signature. Indeed, the simple ballistic model proposed by \citet{Sakai_2014} was meant to explain asymmetric velocity structure in L1527 via parabolic motions arising in a dense infalling envelope. However, we emphasize that an advantage of the STAK disk is its reasonable physics, that it connects gas flow from the envelope to the disk by implementing shock physics in a realistic way. 

Additional C$^{18}$O(2-1) ALMA data published by \citet{vanthoff_2018} were meant to look at closer-to-disk scales. However, the velocity moment 1 maps are highly sensitive to the signal-to-noise level and phase errors, and this dataset was too noisy for the particular features we wanted to compare with. Using newer, $0.17\arcsec \times 0.14\arcsec$ high spatial resolution data from the eDisk survey,  \citet{vanthoff_2023} present moment maps for C$^{18}$O(2-1) and other molecular species that remain consistent with the predicted STAK signature. The data have 22~au spatial resolution with a suggested $\sim 100$ au disk radius; our corresponding model in Figure 9 would have 15~au spatial resolution for 75~au disk radius, which lies in-between the 8~au and 25~au model resolutions displayed for partially resolved disks. Comparison by eye shows the expected asymmetry in peak red and blue shifted gas locations with respect to the disk major axis, as well as the inner VCA twist that traces the system velocity. The agreement is encouraging and suggests further analysis would be useful to more fully compare the results. One caution is that there is no standardized method for constructing velocity moment maps, meaning the comparison may depend on the adopted method.

There remain some inconsistencies with regard to source inclination in L1527, where the spectral line data mostly indicate $i=95^{\circ}$, while the 1.3 mm dust continuum seem to imply $i=85^{\circ}$ instead (\citet{vanthoff_2023} and references therein). The $i=95^{\circ}$ inclination that we favor for C$^{18}$O(2-1) is consistent with earlier analysis by \citet{Oya_2015}. They used a synthesized beam of 0.8$^{\prime\prime}$ $\times$ 0.7$^{\prime\prime}$ and found that CS (J = 5-4) emission that reflects the infalling motion is best reproduced using $i=95^{\circ}$ source inclination. Note that $i=95^{\circ}$ inclination would mean that the rotation axis points to the east of the protostar and away from us  (i.e. Fig. \ref{fig:mom1_ALMA_comparison}, left-hand side of images), so that the eastern side is the nearest side of the disk, namely the side that is closest to the observer. However, the dust continuum emission at 1.3 mm is likely to be optically thick which complicates the interpretation. To resolve the issue, higher spatial observations at optically thin wavelengths are suggested.

In their analysis, \citet{vanthoff_2023} emphasized the difficulty of deriving the disk inclination accurately for the nearly edge-on system. The differences in orientation seem to rely more on the detailed connection in physical structure between the disk and the envelope. These differences may indicate a misalignment between the disk inclination and the envelope orientation or possibly an inner disk warp casting a shadow. In support of the disk warp interpretation, we note that a recent analysis by \citet{Villenave_2024} presents evidence based on JWST near-infrared data that inner $\sim 5$ au disk warps occur frequently for protostars, including L1527. Additional discussion comparing various spectral line and continuum datasets can be found in \citet{sheehan_2022} and \citet{vanthoff_2023}. Furthermore, twisted kinematics in the \cuo{18} velocity map that is seen as S-shape close to the systemic velocity,  are already found in two eDisk sources, the Class 0 GSS30 IRS3 \citep{Santamaria-Miranda_2024}, and the Class I Oph IRS63 \citep{Flores_2023}, suggesting the transition from a Keplerian disk to an infalling envelope.

\subsection{Model advantages and limitations}

The physical structure of our disk is described in the same way as in \citet{Flores-Rivera_2021}, by a 2D density geometry parameterized with a radial power law and vertical Gaussian structure with axial symmetry \citep{Whitney_2013}. As such, the parameterization is a useful tool for comparison with observational data, meant to capture the protostellar system at a snapshot in time. 

For typical disks the vertical disk structure is based on a picture where gas is vertically supported by gas pressure, which if the disk gas is cold ($c_s/v_{c,kep} << 1$ and $v_{c,kep}=(GM/r_{c,kep})^{1/2}$) leads to geometrically thin disks. Similarly, we consider whether the vertical velocity $v_z$ in the STAK disk is small, which would suggest consistency with vertical gas pressure support. Along a disk elliptical streamline, we note that the initial post shock value is $v_z=0$ at $r_{shock}$, with gas then accelerating inwards due to gravity and reaching maximum velocity at the second shock $r_{c,kep}$ in the disk midplane.  To evaluate whether $v_z/(v_{c,kep})$ is small we use expressions from \S \ref{sec:elliptical_streamlines} for the elliptical streamline, specifically Equations \ref{eqn_e}, \ref{eqn_rshock}, \ref{eq13} -\ref{eq15} to find that $v_z/(v_{c,kep}) = \cos{\theta_1}\sin{\theta_0}$. For the fiducial streamline (Table \ref{tab:fiducial_parameters}) the ratio is 0.10, and is similarly small for other streamlines, thus showing consistency with a vertically pressure supported disk.

In this work we update the velocity model to follow the gas motion through the shock that defines the boundary between envelope and disk. However, one limitation is that we do not update the disk density in a self consistent manner, leaving that to future work. 

To place this work in context, we implicitly envision the STAK motions as describing fast/dynamical flow through the surface layers of the disk, before reaching the second shock in the disk midplane, where the gas joins a more massive region of ``settled" material in the disk midplane, having near-circular motion. This structure would be similar to \citet{Harsono_2011}, in their numerical simulation of massive disks with infall. They found that infall leads to sub-Keplerian flow in the disk upper layers, that transitions to circular motion in a roughly hydrostatic disk within one disk scale height of the midplane. Future improvement to the STAK model could include the circular motion of a midplane settled disk. This may shed light on observational analysis of P-V diagrams that show evidence for a radial break in the disk velocity profile \citep{Oya_2015}.  

An additional assumption is that we do not consider shock heating, which we have argued does not strongly affect gas dynamics, but nonetheless is likely to have consequences for dust grain heating and chemical abundances in the uppermost disk layers. In addition, we assume the density distribution is axisymmetric, which is only approximate for real objects. However, including streamers of high density and infalling material, as has been suggested for some objects \citep{delaVillarmois_2022,Yen_2017}, are a straightforward implementation for future modeling. 

We point out that the STAK dynamic disk framework can be extended from what is considered here to include additional physics within the disk. For example, we neglect gas pressure and turbulence in the disk but their effects can be included. Also, we simplified the ram pressure boundary condition to require that all inward motion be quashed at the first shock; however, other assumptions can be considered. Such extensions would therefore modify the orbit away from purely elliptical, namely away from a ballistic elliptical orbit where apoapsis occurs at the first shock.

Our focus on a snapshot in time also means that the STAK  framework can reveal disk structure but in this formulation is not well suited for study of time dependent disk evolution. We have limited our description of the infalling rotating cloud core and disk such that the disk does not include gas pressure, magnetic field, turbulence, or viscosity, all of which may influence disk evolution.

\subsection{Other work}

Several relevant studies of protostars include mass infall and additional physics to what we present. In their theoretical study, \citet{Jones_2022} conducted HD global simulations of a protoplanetary disk accounting for rotation, self-gravity, and viscosity to find the disk edge (maximum cylindrical radius extent), where in principle the midplane gas in the infalling cloud stops when it meets the disk, happens roughly at the centrifugal radius. \citet{Mori_2024} consider, in addition to the UCM envelope, the ballistic model within the context of midplane-only flow (SB model), where material flows inward to a shock located at roughly 0.5 times centrifugal radius, to construct P-V diagrams and assess observational uncertainties in derved parameters such as protostar mass. To model the protostar L1527, \citet{Shariff_2022} perform a 1D time-dependent disk study that includes shocks, radiative and other heating and cooling sources, but no turbulent viscosity. The 1D vertical integration procedure assumes that incoming mass and momenta are instantaneously mixed vertically into the disk. They find that a radial infall shock occurs at roughly 1.5 times the centrifugal radius, and at smaller radii the infalling material piles up near the centrifugal radius. In consequence there is a difference between the disk edge as determined by mass surface density, and disk edge as determined by velocity due to an infall shock. In their MHD study \citet{Hennebelle_2020} include rotation, turbulence, and misalignment between rotation and magnetic field axis to show that disks of suitable size but relatively low-mass are able to form. In contrast, \citet{Harsono_2011} perform numerical simulation of massive disks with gravitational instability, and find that infall leads to sub-Keplerian flow in the disk upper layers, that transitions to circular motion within one disk scale height of the midplane. However, the spiral features present in the simulations by \citet{Hennebelle_2020} and \citet{Harsono_2011} and that transport angular momentum are not visible in high fidelity disk observations, specifically the eDisk continuum survey of protostar disks \citep{Ohashi_2023}. This may be due to contrast or optical depth effects; longer wavelength data could help resolve this issue.

\subsection{Other mechanisms}

The extent to which Class 0 or I disks involve elliptical motion for the gas is unknown and merits further investigation. In particular, later stage Class I disks experience lower mass infall rates, such that the STAK signature may be reduced or disappear relative to an established settled disk. For example, the IRAS 04302+2247 is a late stage Class I source that is a highly inclined (87$^{\circ}$) edge-on disk and from the dust continuum and gas kinematics suggest that the disk is strongly settling dust into the midplane \citep{Lin_2023}. The molecular line emission, having 16~au resolution, shows a characteristic butterfly pattern in the moment 0 and moment 1 maps suggesting that the gas follows circular Keplerian motion. However, the data also show evidence for rotating envelope material along with a hint of the twist angle for the elliptical STAK signature, suggesting that additional modeling may be useful. An additional avenue to investigate are the presence of non-axisymmetrical structures such as large-scale streamers that can cause accretion shocks in the outer disk that could also be affecting the rotation axis of the disk-envelope \citep{delaVillarmois_2022}. Further 3D physical modeling must be done to analyze the relation between the mass of the envelope and the angular momentum vector of the infalling material. The lack of alignment between the envelope's rotational axis and the initial direction of the magnetic field can lead to a distorted disk structure as the protostellar core collapses as demonstrated by \citet{Hirano_2020}. Additionally, the magnetic fields within protostellar cores show a seemingly stochastic orientation in relation to their associated outflows \citep{Hull_2014, Lee_2017}. Therefore each system must be carefully studied individually. 

There are relatively few young ($<1$ Myrs) Class II disks that show evidence of envelope remnants. The iconic HL Tau is late stage I or early stage II, and its continuum emission shows a $\sim 100$~au radius disk containing rings and gaps indicating possible planet formation \citep{ALMAPartnership_2015}. However, the molecular line emission has significantly lower spatial resolution \citep{Wu_2018,Yen_2017}, comparable to the situation shown in Figure \ref{fig:mom1_ALMA_comparison} for L1527. There is a hint of the STAK twist in the velocity moment 1 map, which suggests a need for higher spatial resolution data to better resolve the motion. The AB Aurigae system shows evidence of inhomogeneous accretion from the remnant envelope as well as some outer parts in CO that are not consistent with the disk Keplerian rotation \citep{Tang_2012, Pietu_2005}. RU Lup also shows a non-Keplerian envelope-like structure in CO with some spiral arms stretching up to beyond 200 au \citep{Huang_2020}. RU Lup is also undergoing sudden increases in brightness \citep[i.e.,][]{Giovannelli_1991} which points to outbursting FU Orionis system category that can be related to gravitational instability \citep[GI;][]{Boss_1997,Gammie_2001}. Another candidate for GI that also exhibits infall motions from the remnant envelope is Elias-227 \citep{Paneque_2021}. 

For other Class II ($>1$ Myrs) disks with no evidence of envelope remnants, their gas kinematics are instead shaped by the disk viscous evolution, forming-planets, flyby perturbers, or by some dispersal mechanism (i.e., MHD winds). For CQ Tau, the peak intensity map \citep{Wolfer_2021} and the velocity map \citep{Wolfer_2023} shows very similar gas structure comparable to our elliptical modeling case. At the same time, this system demonstrate complex kinematics and physical structures making it complicated to explain them from a single mechanism. Other possible scenarios are an inner disk misalignment caused by a giant planet in the innermost regions.

\section{Conclusions} \label{sec:conclusions}

The gravitational collapse of a rotating cloud leads to mass infall that feeds a gaseous accretion disk and over time builds a central protostar.  Here we focus on how to include shocks with reasonable physics to investigate gas dynamics across the shock, where the infalling cloud meets the outer regions of the disk.  Our theoretical and modeling framework, the shock twist-angle Keplerian or STAK disk, specifies the shock location and the abrupt change in direction of gas flowing across the shock and its subsequent flow through the disk.  We implement the STAK model by updating \citet{Flores-Rivera_2021} so infalling gas streamlines at different polar angles approach and intercept the disk's surface subject to an updated ram-pressure boundary condition.  By considering the energy dissipation and angular momentum conservation across the shock, we argue that gas parcels transition from parabolic free-fall orbits outside to lower-energy elliptical orbits inside the disk.  The resulting disk gas streamlines deviate from circular Keplerian orbits in ways detectable using spectral line data from interferometers such as ALMA.  We adopt the same physical and chemical structure and parameters presented in \citet{Flores-Rivera_2021} to produce synthetic intensity and velocity maps to examine the differences between disks on circular and elliptical orbits.  Our key findings are summarized as follows:

\begin{itemize}[label={--}]
    \item The STAK disk exhibits a distinctive signature where the intensity and velocity moment maps show an inner twist with respect to the underlying disk structure traced by the dust continuum, that arises due to the abrupt change in gas velocity direction across the shock.  This observable inner twist deviates from the disk major axis, contrary to the circular Keplerian case. The velocity signature shows some similarity to that of parabolic orbits arising from pure infalling envelope motion. However, the disk physics of post-shock gas with rapid cooling dictates that the disk naturally has higher density and thus dominates the model signature.
    \item The STAK disk's most distinctive features vary with the source inclination and telescope angular resolution.  For disks that are only marginally spatially-resolved, the STAK signature is retained best in the velocity moment map.
    \item  We implement a STAK disk model for the protostar L1527 using the RadChemT code, and find improved agreement with archival ALMA data when compared with the circular Keplerian model.  More recent high-spatial-resolution ALMA data also appear consistent with the predicted inner twist.  Furthermore, hints of the STAK disk signature appear in published ALMA data for other protostars.
    \item We encourage future spectral line observations that fully resolve the disk to search for the kinematic signatures produced by the envelope-disk transition.  We propose characterizing the velocity asymmetry in the moment map by tracing a line at the systemic velocity using what we call the Velocity Crowding Angle, defined in \S \ref{sec:velocity_sig}, Fig.~\ref{fig:moment1_maps}, and Fig.~\ref{fig:spatial_res_mom1}.
    \item  The C$^{18}$O gas motion of L1527 in ALMA archival data at both envelope and disk scales best matches a source inclination of 95$^{\circ}$, rather than 85$^{\circ}$. This is defined using the right hand rule for the rotation axis, measured relative to the observer's line of sight. The STAK disk kinematics resolve the ambiguity over which side of the disk is closer to the observer. 
 
\end{itemize}

The theoretical framework we present for the envelope-disk shock is semi-analytic and thus readily implemented to fit observational data over the large available parameter space. It builds on existing use of Monte Carlo radiative transfer modeling codes that compute temperature from the specified protostar luminosity and density distribution, but includes adding physically realistic velocity distributions to connect the infalling envelope to subsequent gas motion within the disk. The framework can be used to enhance current snapshot-in-time fitting, when fitting with numerical hydrodynamical simulations is computationally infeasible. 

The physical model adopted in the current work assumes a geometrically thin disk, with the centrifugal radius defining the disk radius of both the gas and dust. However, observations show cases where the gas disk radius appears larger than that of the dust. Future directions that merit investigation are to generalize the disk model, and also to consider how the gas disk's size might be measured using different dense gas tracers.
   
The STAK approach enables searching existing and new measurements for evidence of the envelope-disk shock.  Many Class 0, I, and young Class II disks are good candidates since they show evidence of gas falling onto their disks.  A future direction will be to adapt the paradigm to model non-axisymmetric infall and treat streamers that may produce localized accretion shocks on the outer disk. 


\section{Acknowledgments}
We dedicate this paper to the memory of Frank H.\ Shu.  S.\ Terebey thanks the Jet Propulsion Laboratory for its hospitality and acknowledges this work was supported in part by NASA grant NNX15AQ06A.  L.\ Sandoval Ascencio acknowledges this work was supported by MORE Programs RISE MS-to-PhD and the National Institute of General Medicinal Sciences, NIH, support through TWD RISE award R25 GM061331.  L.\ Flores-Rivera acknowledges this work was funded by the European Research Council (ERC Starting Grant 101041466-EXODOSS).  N.\ Turner's efforts were supported by NASA's Exoplanets Research Program through grant 17-XRP17\_2-0081.  The research was performed in part at the Jet Propulsion Laboratory, California Institute of Technology, under contract 80NM0018D0004 with the National Aeronautics and Space Administration.

%

\vspace{5mm}
\facilities{ALMA}


\software{astropy}



\appendix

\section{Effective potential} \label{app:veff}
We derive an expression for the effective potential. In spherical coordinates the mechanical energy E per unit mass (kinetic plus potential) is given by,
\begin{equation}
E = \frac{1}{2} (v_r^2 + v_{\theta}^2 + v_{\phi}^2) - \frac{GM}{r}, 
\end{equation}
which can be rearranged to give,
\begin{equation}
  E = \frac{1}{2}(v_r^2 + v_{\theta}^2) + \frac{v_{\phi}^2r^2\sin^2(\theta)}{2r^2\sin^2(\theta)} - \frac{GM}{r}.
\end{equation}

For a central force the angular momentum per unit mass $\bf{\Gamma_n=r\times v}$ is conserved. In an axisymmetric system the cylindrical component $\Gamma =\Gamma_n \sin\theta_{0}$, aligned with the rotation axis of the cloud, is also constant. Using the cylindrical radius $R = r \sin\theta$ then $\Gamma = R v_{\phi} = r \sin\theta v_{\phi} = constant$, leading to 
\begin{equation}
  E = \frac{1}{2}(v_r^2 + v_{\theta}^2) + \frac{\Gamma^2}{2r^2\sin^2(\theta)} - \frac{GM}{r},
\end{equation}
Now define the effective potential $V_{eff}$ to be
\begin{equation}
  V_{eff} =  \frac{\Gamma^2}{2r^2\sin^2(\theta)} - \frac{GM}{r},
\end{equation}
The expression includes both $r$ and $\theta$ coordinates. However, the expression for $V_{eff}$ simplifies when evaluated in the disk midplane at $\theta =\pi/2$, $\sin(\theta)=1$ and cylindrical $R=r\sin(\theta)=r$, to give 
\begin{equation}
  V_{eff} =  \frac{\Gamma^2}{2r^2} - \frac{GM}{r},
\end{equation}
thus giving the expression that is plotted in Figure \ref{fig:effectivepotential}, and for which the same value of $\Gamma$ connects the three orbital segments: parabolic, elliptical, and circular.

\section{Cylindrical angular momentum} \label{app:ang_mom}
Here we describe how constant cylindrical angular momentum $\Gamma$ relates the three orbital segments: parabolic, elliptical, and circular, and derive where the orbit segments cross the disk midplane.

Following \citet{Cassen_1981}, consider next that the angular momentum is separable into two parts, 
\begin{equation}\label{gamma_appA}
\Gamma = \Gamma_{\infty} f(\theta_{0}),
\end{equation}
where the function $f(\theta_{0}$) describes the angular dependence at large $r$ and $\Gamma_{\infty}$ represents the maximum instantaneous angular momentum for the $\theta = 90^{\circ}$ equatorial streamline.
The function $\Gamma_{\infty}=\Gamma_{\infty}(t)$ describes the angular momentum originating at large $r$ that is just now reaching the central region. Moreover, $\Gamma_{\infty}(t)$ describes slowly varying ($\sim 10^5$ years) changes related to the outer cloud core, namely slow compared with disk orbital period. See \citet{Terebey_1984} who demonstrate that the inside-out collapse solution for a slowly rotating cloud has the \citet{Cassen_1981} solution as an inner asymptotic limit: a partial description follows. For solid body rotation then $\Omega=constant$; prior to collapse the cylindrical component of angular momentum is constant on cylinders having cylindrical radius $R = r \sin(\theta_0)$ and  $v_\phi = R \Omega =\Omega r \sin(\theta_0)$ so that $\Gamma = R v_{\phi} = \Omega r^2 \sin^2(\theta_0)$. This expression demonstrates that the function $f(\theta_{0})= \sin^2(\theta_{0})$ for solid body rotation. Further, the mass interior to radius $r$ is given by $M(r) = 2 c_s^2 r/G$ for the initial cloud. One can then use the similarity collapse solution \citep{Shu_1977} to relate collapse age ($t=r/c_s$ defined at the expansion radius) to the enclosed mass, and then to the originating radius (see TSC). Physically, the mass just now reaching the central region (i.e. disk edge) originated at radius $r=(m_0/2) c_s t $, where $m_0= 0.975$ is a constant. Substituting for $r$ in $\Gamma$ leads to the expression $\Gamma_{\infty}(t) =\Omega [m_0 c_s /2]^2 t^2$ for the time dependent behavior.

For a parabolic orbit, substituting equations (\ref{gamma_def}) and (\ref{gamma_appA}) into equation (\ref{ell_def}), and evaluating the radius Equation \ref{eq4} in the disk midplane at $\theta =\pi/2$ and where $r=\ell$ leads to 
\begin{equation}
r=\frac{\Gamma_{\infty}^{2}}{GM} \frac{f^{2}(\theta_{0})}{\sin^{2}(\theta_{0})} ,
\end{equation}
Assuming a pre-collapse cloud with constant angular velocity $\Omega$ , as we do here, then $f(\theta_{0})= \sin^2(\theta_{0})$, leading to an expression for the radius where a parabolic orbit crosses the disk midplane,
\begin{equation}\label{eqn_semi-latus-par}
\ell =\frac{\Gamma_{\infty}^{2}}{GM} {\sin^{2}(\theta_{0})}.
\end{equation}

The parabolic streamline containing the maximum angular momentum has $\theta_{0}= \pi /2$, leading naturally to an instantaneous disk maximum radius, 

\begin{equation}\label{eqn_rd}
r_d=\frac{\Gamma_{\infty}^{2}}{GM} .
\end{equation}
In the literature this radius for the disk is commonly called the centrifugal radius.

The radius $r_\mathrm{c,kep}$ of the destination orbit is simply expressed in terms of $\Gamma$, the cylindrical angular momentum. Rewriting the specific angular momentum in terms of cylindrical radius $ r \sin\theta$ then,
\begin{equation}
\Gamma =r \sin\theta v_{\phi}.
\end{equation}
In the disk midplane, $\Gamma =r v_{\phi}$. Recall that for a circular Keplerian orbit,
\begin{equation}
v^2_{\phi} = GM/r.
\end{equation}
Combining expressions leads to,
\begin{equation}\label{eqn_gamma_circ}
\Gamma^2 =GMr_\mathrm{c,kep}.
\end{equation}

Rewriting in terms of disk radius $r_d$ rather than $\Gamma$ using equations \ref{gamma_appA}, \ref{eqn_rd} and \ref{eqn_gamma_circ} leads to the expression,
\begin{equation}\label{eqn_rcirc}
r_\mathrm{c,kep} = r_d f^2(\theta) = r_d \sin^4(\theta_0),
\end{equation}
which gives the destination circular orbit, in terms of the infalling $\theta_0$ parabolic streamline angles.
We have so far followed the discussion in \citet{Cassen_1981}, who describe the angular momentum in the parabolic orbit of the infalling gas, and who show that the gas enters the disk at sub-Keplerian speeds. However, our assumptions differ at the infall shock, where we assume that infalling envelope gas transitions to an elliptical orbit when it crosses the shock defining the envelope-disk surface. The task at hand is to define the properties of the elliptical orbit, and to justify the conditions under which it has the same angular momentum as does the destination circular orbit. The known quantities are ($r_\mathrm{shock},\theta_1,\phi_1$), which is identified with apoapsis, and $r_\mathrm{c,kep}$, which specifies the angular momentum (equation \ref{eqn_gamma_circ}). As we argue in \S\ \ref{sec:elliptical_streamlines}, a reasonable condition is  constant $v_{\phi}$ across the first shock where envelope material enters the disk, and zero inward poloidal motion for the streamline. With that assumption, then immediately post-shock the disk gas inherits the cylindrical angular momentum of the pre-shock gas, having cylindrical $r_\mathrm{shock}=r_\mathrm{shock} sin(\theta_1)$ so that, 
\begin{equation}\label{eqn_gamma_ellipse}
\Gamma=r_\mathrm{shock} \sin(\theta_1) v_{\phi}, 
\end{equation}
where $v_{\phi}$ is evaluated at the shock location. This location defines the orbital plane of the ellipse and has polar angle $\theta_1$, which differs from $\theta_0$ for the parabolic orbit. 

To continue, the general equation for a conic section (see \citet{Goldstein2002} equations 3.55 and 3.56) can be written using coordinates in the orbital plane as,
\begin{equation}
r' = \frac{C}{(1+e \cos\theta')};  ~~C = {a(1-e^2)}, ~C = \frac{\Gamma_{n}^2}{GM},
\end{equation}
with $\Gamma_{n}$ being the angular momentum normal to the plane of the orbit, and where the shape of the ellipse may be specified in terms of geometrical parameters (a,e) or physical parameters ($E,\Gamma_{n}$) by using the additional relation for energy per unit mass $E = -GM/(2a)$. Using subscript $e$, to distinguish the ellipse from the parabola, then the semi-latus rectum of the ellipse, $\ell_e$, occurs where $\theta' = \pm \pi/2$, for which $cos(\theta')=0$ so that $r'=\ell_e$ and,

\begin{equation}\label{eqn_ell_e}
\ell_e = a(1-e^2) =  \frac{\Gamma_{n,e}^2}{GM}.
\end{equation}

Notably, this location also marks where the ellipse crosses the disk mid-plane, namely $z=0$ and $\theta = \pi/2$ in spherical coordinates. Recalling that cylindrical $\Gamma = \Gamma_{n,e} sin(\theta_1)$, and eliminating $\Gamma$ using equation (\ref{eqn_ell_e}) and equation (\ref{eqn_gamma_circ}) leads to the desired relation between the circular and elliptical orbits, 
\begin{equation}
\ell_e =  \frac {r_\mathrm{c,kep}}{\sin^2(\theta_1)}.
\end{equation}

The disk is assumed geometrically thin, and therefore the polar angle $\theta_1$, that defines the disk surface, satisfies the condition $\sin(\theta_1) \approx 1$. This is illustrated using the fiducial streamline, for which $\theta_1 =83\degr$ (see Table \ref{tab:fiducial_parameters}) for which $\sin^2(\theta_1)=0.985$. Therefore we conclude that,
\begin{equation}\label{eqn_ell_rckep}
\ell_e \approx r_\mathrm{c,kep}.
\end{equation}

Physically, this means that gas flowing through the disk on the elliptical streamline orbit has approximately the correct angular momentum to settle into a circular Keplerian orbit, after going through a second shock when it reaches the disk midplane. This is the destination orbit.

The geometrical parameters (a,e) for the ellipse can now be determined by recalling that apoapsis means $r_\mathrm{shock} = a(1+e)$ and using equations \ref{eqn_ell_e} and \ref{eqn_ell_rckep}  so that eccentricity is determined from, 
\begin{equation}
(1-e) = \frac{r_\mathrm{c,kep}}{r_\mathrm{shock}},
\end{equation}
and semi-major axis from,
\begin{equation}\label{app_eqn_a}
a = \frac{r_\mathrm{shock}}{(1+e)}.
\end{equation}
These parameters specify the shape of the ellipse described in \S\ \ref{sec:elliptical_streamlines}. For our purpose, the velocity along the orbit is also needed. For the Kepler problem, the velocity lies within the orbital plane,
\begin{equation}
\bold{v'_{\parallel}}= v_{r'}\bold{\hat{r'}} + v_{\theta'}\hat{\bold{\theta'}},
\end{equation}
and has a known form (see chapter section 3.7 of \citet{Goldstein2002}), where the velocity components can be expressed in terms of (a,e) such that,
\begin{eqnarray}\label{app_eqn_vcomponents}
v_{r'} = \frac {e v_o \sin\theta'} {1-e^2}, \nonumber \\
v_{\theta'} = \frac{a v_o}{r'} = \frac{a v_o (1+e\cos\theta')}{r_o}.
\end{eqnarray}
In this form, the elliptical velocity components are defined relative to a circular orbit ($r_o,v_o=\sqrt{GM/r_o}$) having the same angular momentum $\Gamma_{n,e}$, and with the specification that $r_o=a$ (for us, using $a$ from equation \ref{app_eqn_a}), so that $v_o=\sqrt{GM/a}$ is also determined. For visualization, recall that the origin of the elliptical orbit equation is described relative to the protostar located at a focus, whereas the reference circular orbit has radius $r_o=a$ with respect to the geometric center of the ellipse. The reference circular orbit should not be confused with $r_\mathrm{c,kep}$, the destination circular orbit that is defined in terms of $\Gamma$, the cylindrical angular momentum.

\newpage

\section{Post-Shock Cooling Timescale} \label{subsec:coolingt}

We simulate a strong shock scenario where gas inflow into the shock region reaches hypersonic velocities. 
Our prediction of a rapid post-shock cooling suggests that cooling predominantly occurs near the shock location. 
Rapid cooling implies pressure is relatively static, justifying our assumption of ballistic trajectories in the disk. 
While the implementation of post-shock temperatures is beyond the scope of this paper, we validate our assertion by calculating a range of cooling timescales for streamlines in the outer half of the L1527's protostellar disk. 
Under this modeling, we consider a disk radius of r$_{d}=75$ au and a protostar mass of $M_{*}=0.22 M_{\odot}$.
This assertion is pivotal due to the strong dependence of astrochemistry on temperature and time in our model. 
At the shock location, we estimate an upstream temperature of approximately T$_{1}$$\sim35$ K at the shock location \citep[see top panel of Fig. 3 in][]{Flores-Rivera_2021}. 
Employing Equation \ref{mach_num}, we compute Mach numbers for our streamlines, assuming cold gas of solar composition ($\gamma=7/5$, $\bar{m}=2.3 m_{H}$) and estimating both the thermal sound speed ($c_{s}$) and free-fall velocity ($v_{ff}$).
Using our $r_\mathrm{shock}$ values estimated from the ram pressure boundary condition (see \S \ref{sec:disk surface}), we calculate $v_{ff}$ and find it to range between $v_{ff} \sim 2.3$-3.2 km s$^{-1}$. 
Shocks modeled in the disk midplane, rather than along the disk surface as in this note, would consist of calculating $v_{ff}$ at the location where parabolic orbits cross the disk midplane, described by Equation \ref{semilatus}. The Mach number is calculated using
\begin{equation}\label{mach_num}
M_{1}=\gamma^{-1/2} v_{ff}/c_{s}. 
\end{equation}
For a perfect gas, the upstream to downstream temperature ratio can be calculated via the the Rankine-Hugoniot jump condition shown in Equation \ref{temp_ratio} \citep{Shu1992}.
\begin{equation}\label{temp_ratio}
\frac{T_2}{T_1}=\frac{[(\gamma + 1) + 2\gamma (M_{1}^{2} - 1)][(\gamma + 1)+(\gamma - 1)(M_{1}^{2} - 1)]}{(\gamma + 1)^{2} M_{1}^{2}}
\end{equation}
For streamlines with $r_\mathrm{shock}$ $>$ r$_{d}$/2, we find the downstream (hence, post-shock) temperatures range from approximately $T_{2}\sim 240$-440 K. 
Finally, the cooling timescale can be calculated from the internal energy per unit mass for a perfect gas given by Equation \ref{eq_internale}. 
\begin{equation}\label{eq_internale}
\varepsilon = \frac{1}{\gamma-1} \frac{P}{\rho}= \frac{1}{\gamma-1}c_{s}^2   
\end{equation}
We supplement the above equation with a cooling term due to attenuated emission from matter \citep{Turner2001}, given by $4\pi \kappa_{p} B$ where $\kappa_{p}$ is the Planck mean opacity estimated from Figure 9 in \citet{Bell1994} and $B=\frac{\sigma T^{4}}{\pi }$ is the Planck function. 
Finally, we use Equation \ref{eq_cooltime} and the estimated Planck mean opacity of $\kappa_{p}=2$ cm$^{2}$/g to find our cooling timescales.
\begin{equation}\label{eq_cooltime}
t_{cooling}= \frac{\varepsilon}{4\pi \kappa_{p} B}= \frac{\varepsilon}{4\kappa_{p} \sigma T_{2}^{4}}
\end{equation}
For streamlines with $r_\mathrm{shock}$ $>$ r$_{d}$/2, our calculated cooling timescales are notably short, spanning approximately 3-40 minutes.

\newpage
\section{Velocity Connections in a Larger Field of View} \label{largeFOV}

\begin{figure*}[h] 
\centering
\includegraphics[width=15cm]{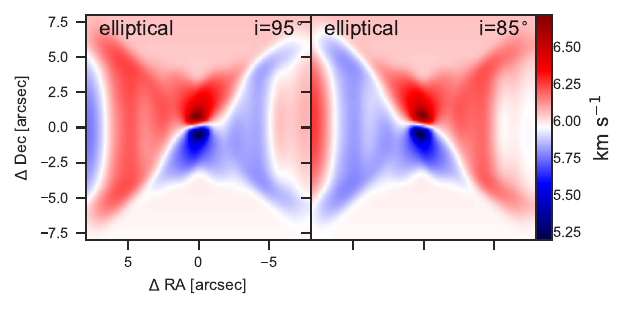} 
\caption{Velocity moment 1 images showing larger scale 2865~au field of view, for two inclinations. L1527 C$^{18}$O elliptical disk model at source inclination 85$^{\circ}$ (left) and 95$^{\circ}$ (right), at low spatial resolution. Near the outflow and away from the disk, the white curve showing the system velocity is seen to change orientation, depending on the assumed source inclination. Color bar shows corresponding velocity in $\kms$. }
\label{outflow_shell}
\end{figure*}

The connection of the disk (north-south vertical) to the larger scale outflow shell (east-west horizontal) in the simulation is clearly visible in Figure \ref{outflow_shell}. The system velocity (curve in white) switches orientation, depending on assumed source inclination for this nearly edge-on system. There are some faint stripes in the simulated outflow shell that arise from a gridding artifact in the modeling, associated with the steep change in density between the outflow shell and the outflow cavity. See \citet{Flores-Rivera_2021} for a comparison with CARMA data showing the CO outflow shell and $3~\kms$ outflow shell velocity.


\bibliography{sample631}{}
\bibliographystyle{aasjournal}



\end{document}